\documentclass[useAMS,usenatbib,usegraphicx]{mn2e}

\title[The evolution of the near-IR LF of galaxies to $z\sim3$]
{Probing the evolution of the near-IR luminosity function of galaxies 
to $z\simeq3$ in the Hubble Deep Field South}
\author[P. Saracco et al.]{P. Saracco$^{1}$\thanks{E-mail:
saracco@brera.mi.astro.it}, A. Fiano$^{2,1}$, G. Chincarini$^{2,1}$, 
E. Vanzella$^{3}$, M. Longhetti$^{1}$,
\newauthor  S. Cristiani$^{3}$ A. Fontana$^{4}$, E. Giallongo$^{4}$,
 M. Nonino$^{3}$  \\
$^{1}$INAF - Osservatorio Astronomico di Brera, Via Brera 28, 20121 Milano, Italy\\
$^{2}$Universit\`a degli Studi di Milano-Bicocca, P.za dell'Ateneo Nuovo 1, 20126 Milano, Italy\\
$^{3}$INAF - Osservatorio Astronomico di Trieste, Via Tiepolo 11, 34131 Trieste, Italy \\
$^{4}$INAF - Osservatorio Astronomico di Roma, Via di Frascati 33, 00040 Monte 
Porzio Catone, Italy \\
}
\begin{document}
%


\date{Accepted December 2005}
\pagerange{\pageref{firstpage}--\pageref{lastpage}} \pubyear{2005}
\maketitle
\label{firstpage}
\begin{abstract}
We present the rest-frame Js-band and Ks-band luminosity function (LF)
of a sample of about 300 galaxies selected in the  HDF-S at Ks$\le23$ (Vega).
We use calibrated photometric redshift together with spectroscopic redshift
for 25\% of the sample.
{ The accuracy reached in the photometric redshift estimate is 0.06 
(rms) and the fraction of outliers is 1\%.}   
We find that the rest-frame Js-band luminosities obtained by
extrapolating the observed Js-band photometry are consistent with
those obtained by extrapolating the photometry in the redder H and Ks bands
closer to the rest-frame Js, at least up to $z\sim2$.
Moreover, we find no significant differences among the luminosities
obtained with different spectral libraries.
Thus, our LF  estimate is not dependent
either on the extrapolation made on the best-fitting template 
or on the library of models used to fit the photometry.
The selected sample has allowed to probe the evolution of the LF in the three
redshift bins [0;0.8), [0.8;1.9) and [1.9;4) centered at the median redshift 
$z_m\simeq[0.6,1.2,3]$ and to probe 
the LF at $z_m\simeq0.6$ down to the unprecedented faint 
luminosities M$_{Js}\simeq-13$ and M$_{Ks}\simeq-14$.
{ We find hints of a  raise of the faint end (M$_{Js}>-17$ 
and M$_{Ks}>-18$)  near-IR LF at $z_m\sim0.6$, raise which
cannot be probed at higher redshift with our sample}.
The values of $\alpha$ we estimate are consistent with the local value
and do not show any trend with redshift.
We do not see evidence of evolution from $z=0$ to $z_m\sim0.6$ suggesting
that the population of local bright galaxies was already formed at $z<0.8$.
On the contrary, we clearly detect an evolution of the LF to 
$z_m\sim1.2$ 
characterized by a brightening of M$^*$  and by a decline of $\phi^*$.
To $z_m\sim1.2$ M$^*$ brightens by about 0.4-0.6 mag  and  $\phi^*$ 
decreases by a factor 2-3.
This trend  persists, even if at a less extent, down to $z_m\sim3$ 
both in the Js-band and in the Ks-band LF.
The decline of the number density of bright galaxies seen at $z>0.8$
suggests that a significant fraction of them increases their stellar mass at 
$1<z<2-3$ and that they underwent a strong evolution in this redshift range.
On the other hand, this implies also that 
a significant fraction of local bright/massive galaxies was
already in place at $z>3$.
Thus, our results suggest that  the assembly of massive galaxies 
is spread over a large redshift range and that the increase of their 
stellar mass has been very efficient also at very high redshift at least for
a fraction of them.
\end{abstract}

\begin{keywords}
Galaxies: evolution -- Galaxies: luminosity function --
             Galaxies: formation.
\end{keywords}

\section{Introduction}
The luminosity function (LF) of galaxies is a fundamental statistical tool
to study the populations of galaxies.
Its dependence on  morphological type, on wavelength and on 
look-back time provides constraints on the evolution of the properties
of the whole population of galaxies, of the populations of various 
morphological types and 
on their contribution to the luminosity density at different wavelength.  
The parameters derived by the best-fitting of the LF, the characteristic
luminosity M$^*$, the slope $\alpha$ and the normalization $\phi^*$, provide
strong constraints to the models of galaxy formation.

The recent estimates of the LF based on local wide surveys in the optical, 
such as the Two Degree Field (2dF; Folkes et al. 1999; Madgwick et al. 2002; 
Norberg et al. 2002) and the Sloan Digital Sky Survey 
(SDSS; Blanton et al. 2003) and in the near-IR, such as the Two Micron All
Sky Survey (2MASS; Kochanek et al. 2001), have provided a comprehensive view
of the local LF for different morphological types and wavelengths.
It is now well established that the LF depends on morphological type,
that the faint-end is increasingly dominated by galaxies with late-type 
morphology and spectra and that their number density increases going to 
lower luminosities.
On the contrary, the bright-end is dominated by early-type galaxies whose 
fraction increases with luminosities (e.g. Marzke et al. 1994, 1998; 
Folkes et al. 1999; Kochanek et al. 2001; Bell et al. 2003).

The studies of the LF based on optically selected redshift surveys 
at $z<1$ have provided the first evidences of a differential evolution of 
galaxies.
Lilly et al. (1995a), using the Canada France Redshift Survey (CFRS; 
Lilly et al. 1995b) show that the rest-frame B-band LF of the blue
population of galaxies brightens by about 1 mag to $z\sim1$ contrary
to the red population which shows a little evolution over the redshift 
range probed.
They used the observed I-band photometry to derive the rest-frame B-band 
luminosities.
Subsequent studies confirmed the different behaviors followed by the
LF of the different populations of galaxies at various redshifts
(e.g. Lin et al. 1997; Liu et al. 1998) and the differential evolution
they underwent (e.g. Bell et al. 2004; Wolf et al. 2003).
At higher redshift ($z>1$) the studies of the LF in the optical rest-frame
confirm the presence of the bi-modality and provide evidences of  
luminosity and of density evolution down to $z\sim2-3$ 
(e.g. Poli et al. 2003; Gabasch et al.2004; Giallongo et al. 2005).

Contrary to the light at UV and at optical wavelengths, the near-IR light
is less affected by dust extinction and by ongoing star formation
e.g. Rix \& Riecke 1993; Kauffmann \& Charlot 1998).
Moreover, since the near-IR light of a galaxy is dominated by the evolved 
population of stars, the near-IR light is weakly dependent on galaxy type 
and it is more related to the stellar mass of the galaxy. 
Therefore, the evolution of the LF of galaxies in the near-IR rest-frame
can provide important clues on the history of the stellar mass assembly
in the Universe rather than on the evolution  of the star formation.

The studies of the K-band luminosity function conducted so far have
shown no or little evolution of the population of galaxies at $z<0.4-0.5$
(e.g. Glazebrook et al. 1995; Feulner et al. 2003; Pozzetti et al. 2003)
with respect to the local population (Kochaneck et al. 2001; Cole et al. 2001;
Glazebrook et al. 1995; Gardner et al. 1997).
On the contrary, evidences of evolution
emerge at $z>0.5$, even if  some discrepancies are present among the results
obtained by the various authors (see e.g. Cowie et al. 1996; 
Pozzetti et al. 2003; Drory et al. 2003; Caputi et al. 2004, 2005; 
Dahlen et al. 2005).
One of the possible reasons of these discrepancies is related to the 
rest-frame K-band luminosities, 
usually extrapolated from the observed K-band photometry making use of 
the best-fitting template which should reproduce the unknown SED of the galaxy.
To minimize the uncertainties, photometry at wavelengths long-wards the 
rest-frame K-band would be required.
Given the obvious difficulties in getting deep mid-IR observations,  
some authors have used the observed K-band magnitudes to construct
 the rest-frame J-band LF 
(e.g. Pozzetti et al. 2003; Bolzonella et al. 2002; Dahlen et al. 2005) 
and to constrain the evolution of the LF in the near-IR.
Another reason could be that deep K-band observations  allowing a good
sampling of the LF at magnitudes fainter than M$^*$ over a wide 
redshift range are not so common.
Consequently  the LF faint-wards of the knee is progressively less
constrained with increasing redshift affecting the LF estimate.  

In this work, we aim at measuring the rest-frame Js-band and Ks-band LF
of galaxies and its evolution to $z\simeq3$ through a sample of about 300
galaxies selected at Ks$\le23$ on the Hubble Deep Field-South (HDF-S).
The extremely deep near-IR observations collected on this field 
coupled with the optical HST data
have allowed us to probe the LF over a wide redshift range and
to sample the faint end of the LF down to unprecedented faint luminosities. 
We use photometric redshift spectroscopically calibrated with a sample
of more than 230 spectroscopic redshift, $\sim80$ of which in the HDF-S.
We pay particular attention to the calculation of the rest-frame 
near-IR luminosities by comparing different methods and libraries of models.
In \S 2 we describe the photometric catalogue and the criteria adopted
to construct the final sample of galaxies used to derive the LF.
In \S 3 we describe the procedure used to obtain calibrated photometric 
redshifts and we present the redshift and color distribution of galaxies.
In \S 4 we discuss the derivation of the rest-frame Js-band and Ks-band
absolute magnitudes and assess the influence of the different estimates
on the LF.
The resulting LF at various redshifts is derived in \S 5.
In \S 6 our LFs are compared with the local estimates and with 
those previously obtained by other authors at comparable redshift
to depict the evolution of the LF of galaxies up to $z\sim3$.
In \S 7 we summarize and discuss the results.    

Throughout this paper, magnitudes are in the Vega system unless explicitly
stated otherwise. We adopt an $\Omega_m=0.3$, $\Omega_\Lambda=0.7$ cosmology
with H$_0=70$ Km s$^{-1}$ Mpc$^{-1}$.

\section{Photometric catalog and sample selection}
 \begin{figure}
\centering
\includegraphics[width=8cm,height=8cm]{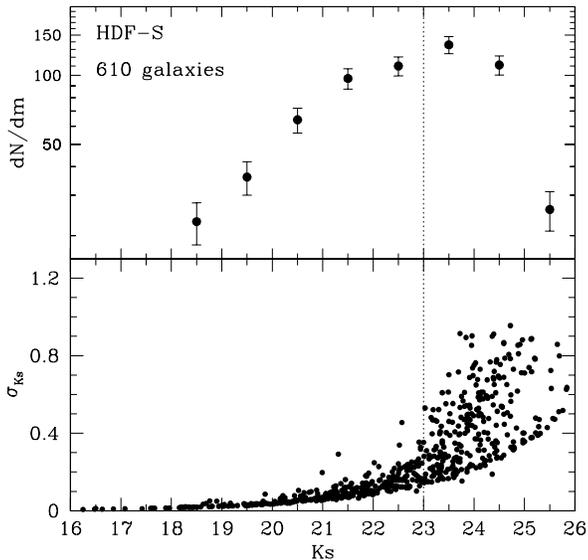}
\caption{Upper panel -- Ks-band number counts   as 
resulting from the whole sample of 610 galaxies detected in the HDF-S.
The counts raise till Ks$\simeq24$  and drop at fainter magnitudes
in agreement with the 100\% completenes estimated at Ks$\simeq23.2$.
Lower panel -- Photometric errors as a function of Ks-band magnitude
for the whole sample of galaxies in the HDF-S. 
At Ks$\simeq23$ the photometric error is still lower than 0.2 mag for
most of the galaxies while is much larger at fainter magnitudes.
The dotted line marks the limiting magnitude of the K23 selected sample.}
\end{figure}
 \begin{figure}
\centering
\includegraphics[width=8cm,height=8cm]{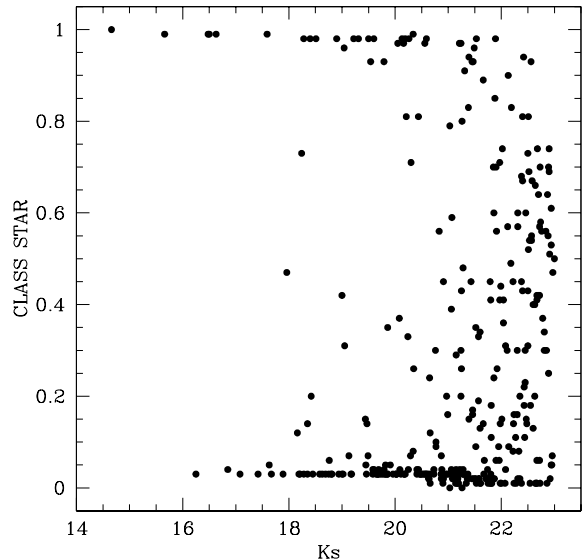}
\caption{Sextractor stellar index ({\small CLASS\_STAR}) as a function of
Ks magnitude for the 332 sources brighter than Ks=23.
}
\end{figure}
The Ks-band selected catalog we use has been obtained by combining
the version 2 of the Hubble Space Telescope (HST) images in the 
U$_{300}$, B$_{450}$, V$_{606}$ and I$_{814}$ bands  (Casertano et al. 2000)
with the deep  ($\sim30$ hrs per filter) 
VLT-ISAAC observations centered on the HDF-S carried out
in the framework of the FIRES project (Franx et al. 2000; Labb\`e et al. 2003a)
 in the Js, H and Ks bands.
{ The formal limiting magnitudes (3$\sigma$ within 0.2 arcsec$^2$) 
reached are 26.7, 29.1 29.5, 28.6, 26.2, 24.5 and 24.4
in  the U$_{300}$, B$_{450}$, V$_{606}$, I$_{814}$, Js, H and Ks bands 
respectively.}
The detection has been performed using SExtractor (Bertin \& Arnouts 1996)
on the Ks-band image.
The Ks-band magnitude is the MAG\_BEST measured by SExtractor while
the colors have been measured within the Ks-band detection isophote
by using SExtractor in double image mode.  
The near-IR data reduction, the detection and the magnitude
estimate are described in detail in Saracco et al. (2001; 2004).
To construct our catalog, we considered the area of the HDF-S 
fully covered by all the images in
the various bands resulting in an area of $\sim5.5$ arcmin$^2$.
On this area we detected 610 sources at Ks$<26$. 
In Fig. 1 the observed number counts are shown. 

From this Ks-band selected  catalog we extracted  a complete sample of 
galaxies  to derive the LF.
On the basis of the simulations described in Saracco et al. (2001), we
estimated that our catalogue is 100\% complete   at Ks$\simeq23.2$.
This is consistent with the number counts shown in Fig. 1 which raise 
till Ks$\simeq24$. 
Down to Ks$\simeq23$ the near-IR photometric errors 
are lower than 0.2 magnitudes while they rapidly increase
at fainter magnitudes as shown in the lower panel of Fig. 1
making very uncertain the photometric redshift estimate.
Thus, according to these considerations, we constructed the sample by  
selecting all the sources (332) brighter than  Ks=23 (K23 sample hereafter). 
We then identified and removed the stars from the K23 sample relying
both on the SExtractor star/galaxy classifier and on the colors
of the sources.
In Fig. 2 the stellar index ({\small CLASS\_STAR}) computed by SExtractor 
for the 332 sources of the K23 sample is shown as a function of the Ks-band 
apparent magnitude.
For magnitudes brighter than Ks$\simeq21$ point like sources
({\small CLASS\_STAR}$\simeq1$) and extended sources 
({\small CLASS\_STAR}$\simeq0$) are well 
segregated while they tend to mix at fainter magnitudes.
We  defined stellar candidates  those sources with magnitudes Ks$<21$ 
having a {\small CLASS\_STAR}$>0.95$ both in the Ks-band image and in the 
I$_{814}$-band image.
We  obtained a sample of 23 stellar candidates.
{ Among them, 7 bright (Ks$<18.5$) sources display spikes in the HST-WFPC 
images implying that they are most likely stars.}
For the remaining 16 candidates we verified that their optical and near-IR 
colors were compatible with those of stars. 
In Fig. 3 the colors Js-Ks and V$_{606}$-I$_{814}$ of the K23 sample 
are plotted as a function of the Ks magnitude and of the Js-Ks color 
respectively.
The 16  star candidates  and the 7 bright stars are marked by filled points 
and starred symbols respectively.
As can be seen, all but 2 of the candidates occupy  the
stellar locus at J-K$<0.9$ and are well segregated both in the 
color-magnitude  and in the color-color plane.
On the contrary, the two candidates lying far from the stellar locus of Fig. 3 
display colors not compatible with those of stars.
In particular, while their J-K color (J-K$>1.3$) could be consistent with 
the near-IR color of an M6-M8 spectral type, the other optical colors differ 
even more than 2 magnitudes from those characterizing these stars.  
Thus, these two sources are most probably misclassified compact galaxies.   
This has made us to classify  21 stars out of the 23 candidates originally 
selected on the basis of the SExtractor stellar index at Ks$<21$.

At magnitudes fainter than Ks=21 other sources (18) lie in the 
locus occupied by stars defined by a color J-K$<0.9$.
By a visual inspection of the 18 sources on the HST  images we clearly 
identified the extended profile for 7 of them. 
On the contrary, the remaining 11 sources have  FWHM and  surface brightness 
profile consistent with a point-like source. 
We probed the stellar nature of these 11 star candidates by comparing their  
observed spectral energy distribution (SED) defined by broad-band photometry 
with a set of spectral templates of stellar atmospheres from the Kurucz's 
atlas (Kurucz 1993{\footnote{www.stsci.edu/hst/observatory/cdbs/k93models.html}}).
The observed SEDs of all the 11 star candidates are very well fitted by 
those of M and K stars.
All but one are not targets of spectroscopic observations, as we verified
through the ESO archive data available.
The spectrum available for the remaining one (J32m58.07s-32'58.9')
confirms its stellar nature as M star.
{ In the left panel of Fig. 4 we show the reduced $\chi^2$ of the best 
fitting template to the SED of the 18 sources with J-K$<0.9$ obtained with 
the stellar atmosphere templates of Kurucz ($\tilde{\chi}^2_{star}$) and with 
the stellar population templates of BC03 ($\tilde{\chi}^2_{gal}$).
It is interesting to note the effectiveness of this method in identifying
stars and galaxies among faint K-selected objects.} 
In the right panel of Fig. 4 we show, as example, the SEDs of 5 out of 
the 11 stars 
together with the relevant best-fitting Kurucz's templates.
Therefore, we classified stars also these 11 sources and
we removed a total of 32 stars from the K23 sample which results in 300 
galaxies.
\begin{figure*}
\centering
\includegraphics[width=8cm,height=8cm]{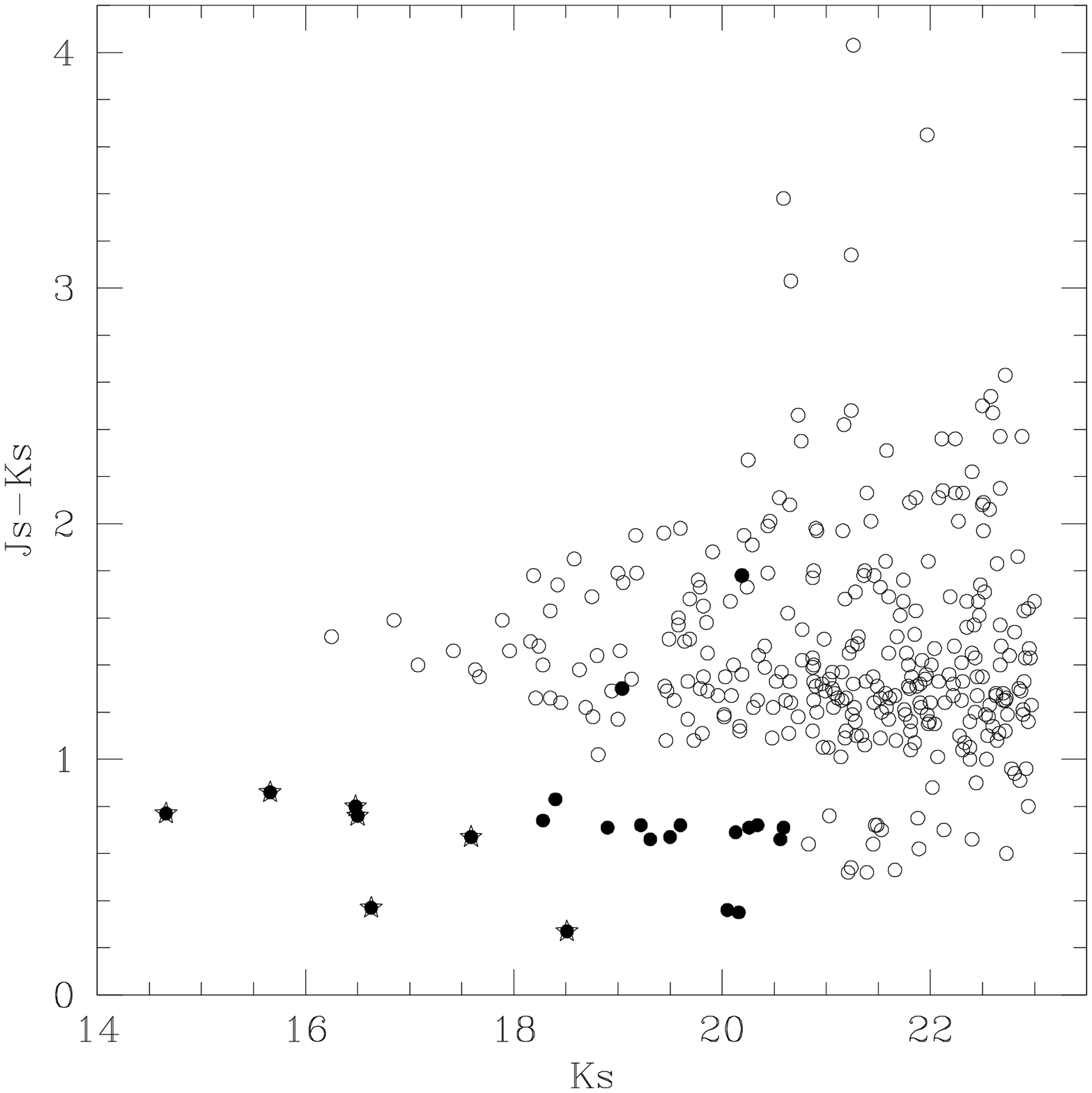}
\includegraphics[width=8cm,height=8cm]{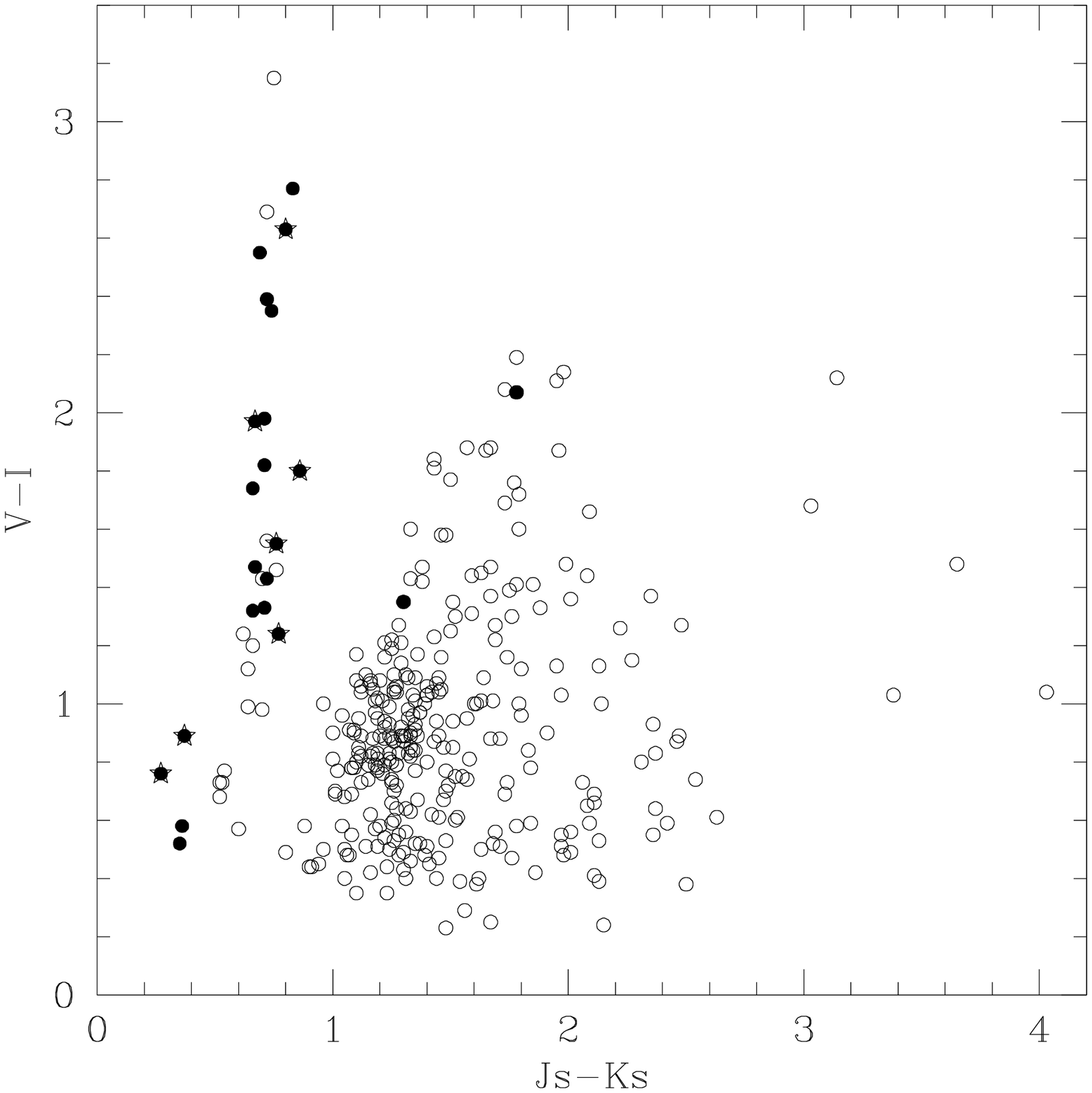}
\caption{Colors Js-Ks (left) and V$_{606}$-I$_{814}$ (right) of the 332
Ks$\le23$ as a function of the Ks magnitude and of the Js-Ks color 
respectively.
The 25 star candidates are marked as filled symbols.
Among them, we marked with starred symbols those sources clearly
identified as stars because showing spikes in the HDF-WFPC images. 
}
\end{figure*}

 \begin{figure*}
\centering
\includegraphics[width=8cm,height=8cm]{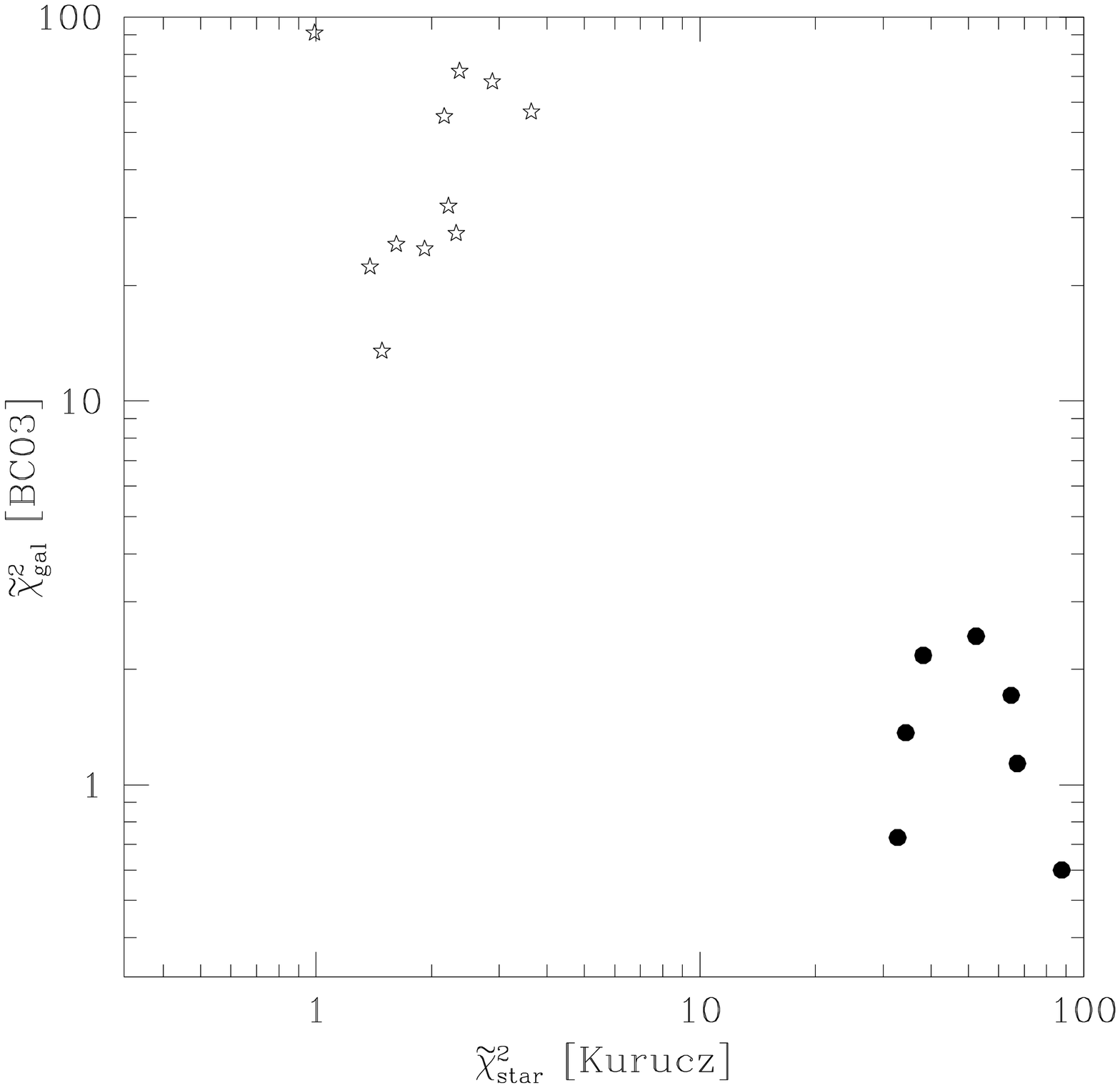}
\includegraphics[width=8cm,height=8cm]{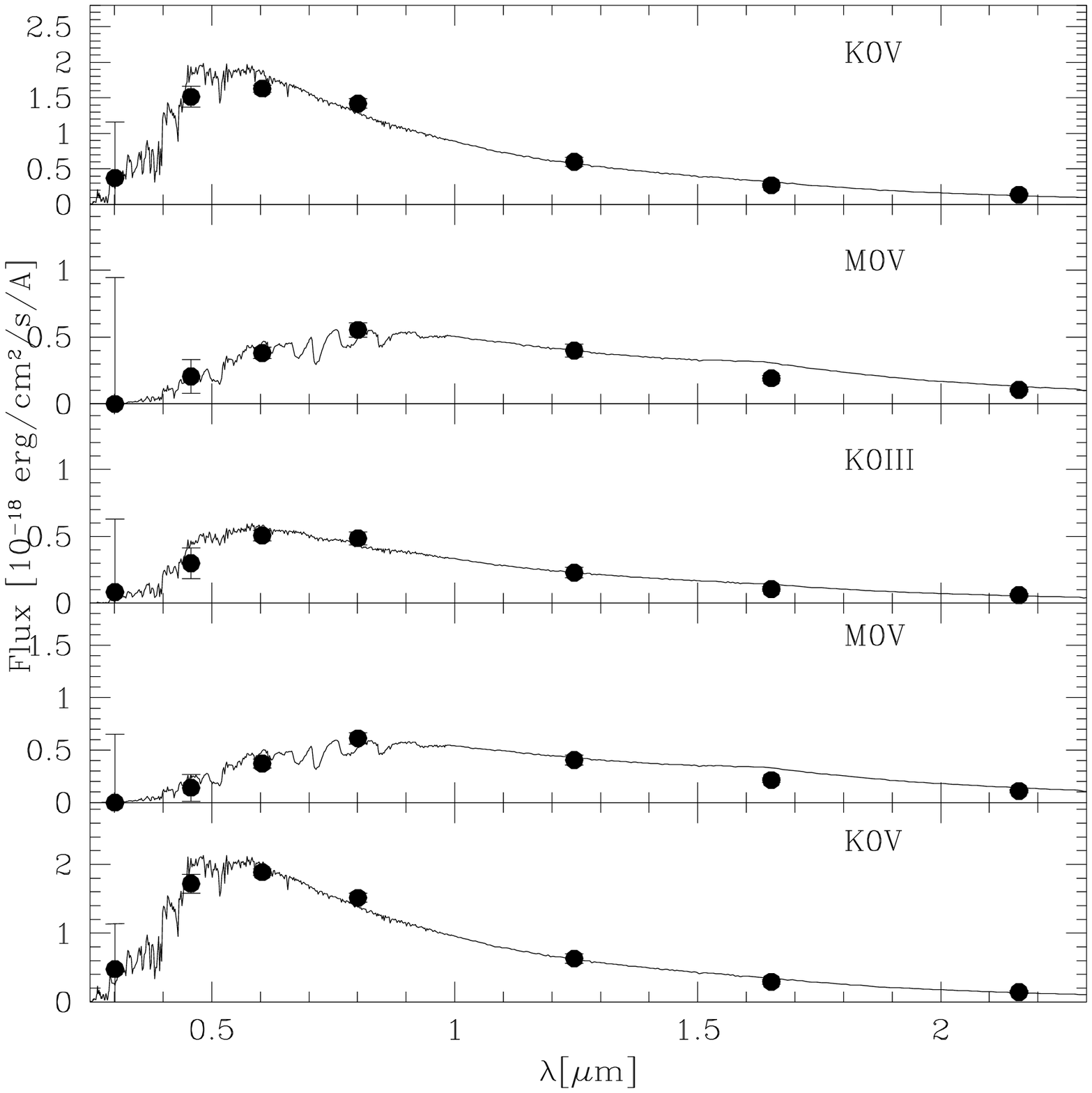}
\caption{{\em Left} - Reduced $\chi^2$ obtained by fitting the SED of the 
18 sources with K$>21$ and J-K$<0.9$ with the Kurucz's templates 
($\tilde{\chi}^2_{star}$) and with stellar population templates 
($\tilde{\chi}^2_{gal}$). The filled symbols mark the 7 galaxies.   
{\em Right} - The templates of stellar atmospheres from the Kurucz's atlas 
are superimposed on the observed SEDs of 5 out of the 11 star candidates 
fainter than Ks$>21$.}
\end{figure*}

\section{Photometric redshifts}
\subsection{Templates selection}
Photometric redshifts have been derived by comparing the observed 
flux densities to a set of synthetic templates based on the latest 
version of the Bruzual \& Charlot (2003) models.
The $\chi^2$ minimization procedure of Hyperz (Bolzonella et al. 2000)
has been used to obtain the best fitting templates.
The set of templates used has been selected among a large  grid of 
models  to
provide us with the most accurate estimate  of the redshift for the
galaxies in our photometric sample.
The grid from which we selected the final set of templates,
includes declining SFRs with time scale $\tau$ in the 
range 0.1-15 Gyr besides the Simple Stellar Population 
(SSP) and the Constant Star Formation (cst) models.
The templates have been produced  at solar metallicity with 
Salpeter, Scalo and Miller-Scalo initial mass functions (IMFs).
Various ranges of extinction A$_V$ and two different extinction laws 
(Calzetti et al. 2000 and Prevot et al. 1984) have been considered in the best-fitting procedure.

The selection of the best set of templates has been performed by comparing 
the photometric redshifts $z_p$ provided by the various sets of templates 
obtainable  by  the whole grid of models, with the spectroscopic redshifts 
$z_s$ of a control sample of  232 galaxies.
This sample includes 151 galaxies in the HDF-N (Cohen et al. 2000; 
Dawson et al. 2001; Fern\'andez-Soto et al. 2002) and 81 in the HDF-S 
(Vanzella et al. 2002; Sawicki \& Mall\`en-Ornelas 2003; Labb\'e et al. 2003b;
Trujillo et al. 2004;
Rigopoulou et al. 2005) with known spectroscopic redshift.
The best set of templates is the one minimizing the mean and the standard 
deviation of the residuals defined as
$\Delta z=({z_{p}- z_{s}})/(1+z_{s})$ and the number of outliers
defined as those sources with $\Delta z> 4\sigma_{\Delta_z}$.
Both mean and standard deviation are iteratively computed applying
a $4\sigma$ clipping till the convergence.
The final number of outliers includes all the galaxies removed by the
clipping procedure.
\begin{table}
\centering
\caption{Parameters defining the best set of templates obtained 
by comparing the photometric redshift with the spectroscopic redshift 
of 232 galaxies, 151 in the HDF-N and 81 in the HDF-S.}
\begin{tabular}{l c}
\hline
\hline
Best set & \\
\hline
SFHs $\tau$ [Gyr] & 0.1, 0.3, 1, 3, 15, cst\\
IMF & Miller-Scalo\\
Metallicity & $Z_{\odot}$\\
Extinction law & Calzetti\\
Extinction & $0\le E(B-V)\le 0.3$ \\
\hline
\end{tabular}
\end{table}

\begin{figure*}
\centering
\includegraphics[width=8cm,height=8cm]{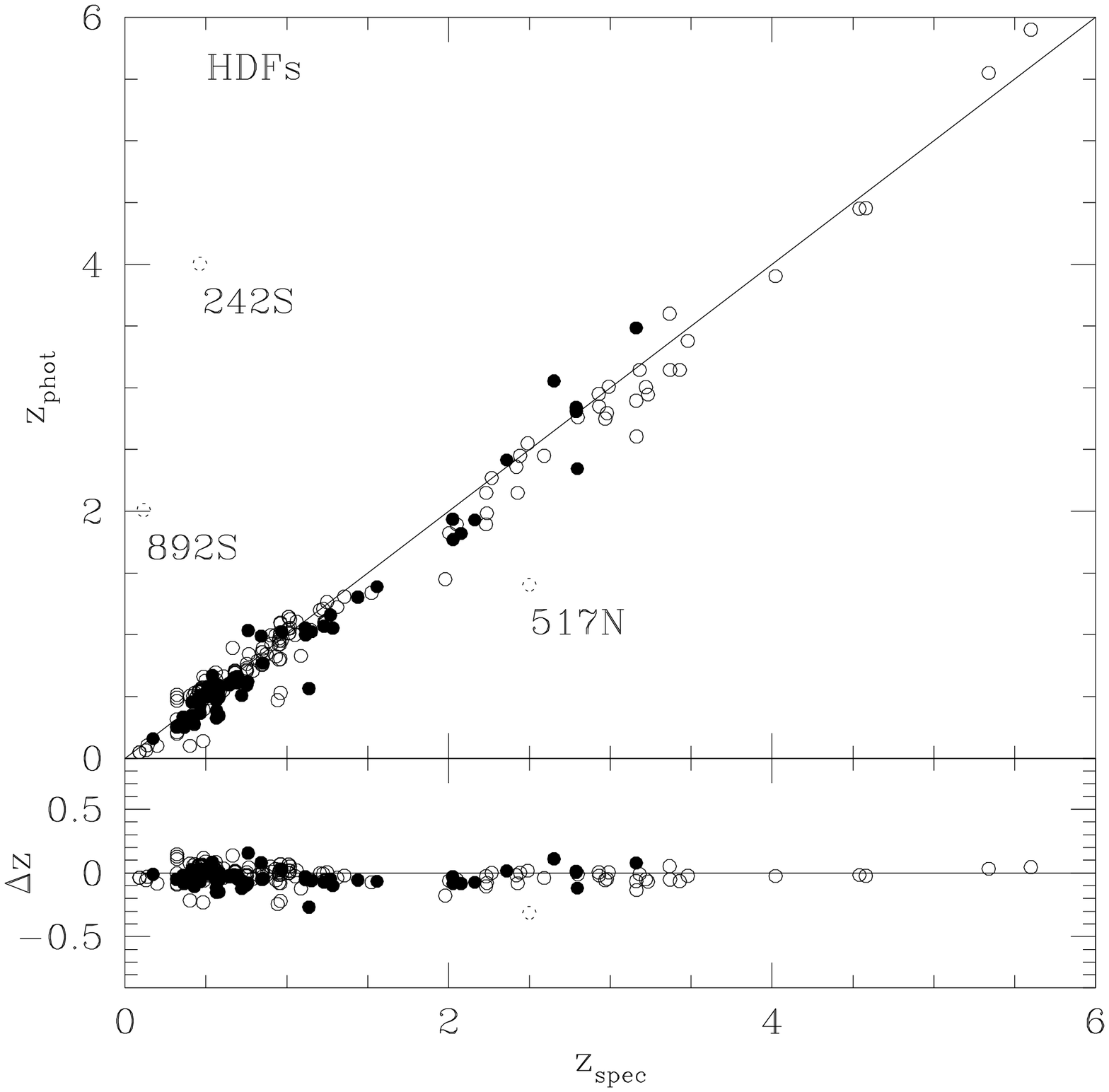}
\includegraphics[width=8cm,height=8cm]{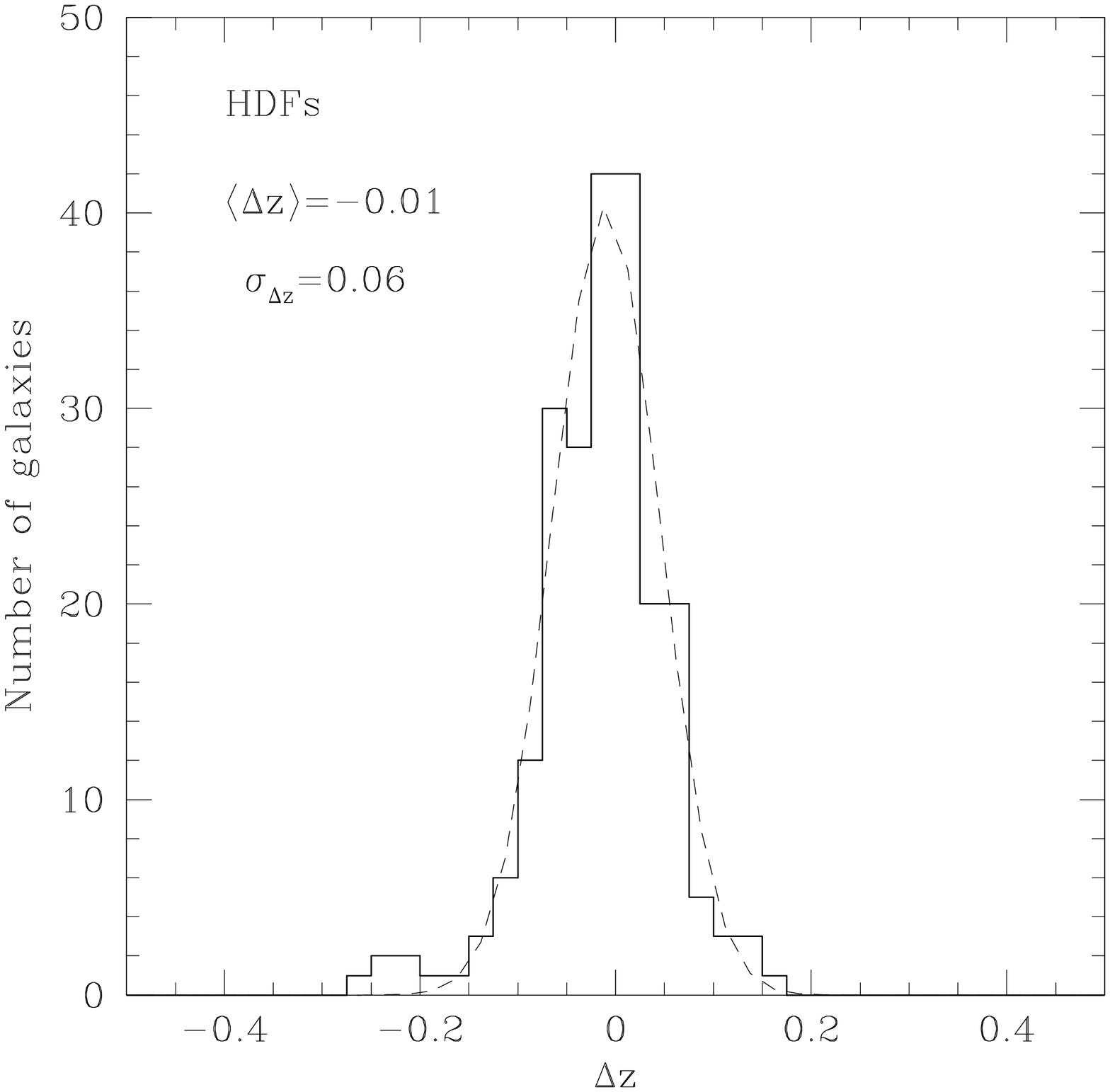}
\caption{-- {\em Left}: Comparison of photometric and spectroscopic redshifts 
for 232 galaxies (upper panel), 151 of which in the HDF-N (empty circles) 
and 81 in the HDF-S (filled circles). 
The outliers are marked by  dotted circles.
The numbers refer to the ID number in the Fernandez-Soto et al. photometric
catalog for the HDF-N and to our photometric catalog for the HDF-S. 
In the lower panel the residuals $\Delta z$ are plotted as a function of the
spectroscopic redshifts. 
 {\em Right} Distribution of the residuals  $\Delta z$ (histogram) for the 
whole spectroscopic sample. A Gaussian (dashed line) with $\sigma=0.065$ and 
a mean deviation of $\langle\Delta z\rangle=-0.015$ from zero is superimposed 
on the observed distribution.   
}
\end{figure*} 
The set of templates and the range of extinction  
providing the best results are summarized in Tab. 1.
It is worth noting that the resulting best set does not include the SSP,
contrary to some of the previous photometric redshift analysis.
This is due to the fact that SSP templates always provide a better fit
to the photometry then other SFHs but a wrong redshift estimate, typically 
lower than the real one.
This results both in a larger deviation of the residuals from the 
null value and in a larger scatter.
In Fig. 5 the photometric redshifts obtained with the best set of templates
are compared to the spectroscopic redshift of the 232 galaxies of our
control sample.
The typical scatter in our photometric redshift estimate is 
$\sigma_{\Delta z}=0.065$ with a  negligible deviation of the residuals 
 from the null value ($\langle\Delta z\rangle=-0.015$).
It can be seen from Fig. 5 that no correlation is present between the
residuals and the redshift. 
We obtain three outliers (1\%) over the redshift range $0<z<6$,
2 out of which (IDs 892 and 242) in the HDF-S.
Both these  are well above the background on the I$_{814}$-band
image while their surface brightness appear very faint in the Ks-band image.
This fact could have implied a wrong estimate of the colors and, 
consequently, of the photometric redshift.
However, looking at the spectrum of 892, we verified that its  
spectroscopic redshift is rather uncertain
since it is based on a single emission line tentatively identified 
as H$\alpha$.
It should be noted that both of them are  fainter than Ks=23,
the magnitude at which  we selected the complete sample to derive the LF
(see \S4 and \S5).
In Tab. 2, we report the values of $\langle\Delta z\rangle$ and 
of $\sigma_{\Delta z}$ together with the number of outliers relevant to the
spectroscopic samples in the HDF-N, in the HDF-S and in the whole sample 
for different redshift bins.

We verified the reliability of our photometric redshifts by comparing 
the spectroscopic and the photometric redshift distributions 
of the 232 galaxies.
In Fig. 6 the cumulative distributions of the photometric and the 
spectroscopic redshift are shown. 
A Kolmogorov-Smirnov test has provided a probability 
P($_{KS}$)$\simeq40\%$ that the two distributions belong to
the same parent population confirming the reliability of our
estimate.
The same result has been obtained considering separately the 151 galaxies
in the HDF-N and the 81 in the HDF-S.
\begin{table*}
\centering
\caption{Mean and standard deviation of the residuals $\Delta z$
for the spectroscopic samples of 232 galaxies (151 galaxies in the HDF-N
and 81 in the HDF-S). For each estimate the results obtained without and
with the 4$\sigma$ clipping are shown.
} 
\label{DzHDFs}
\begin{tabular}{lrlrrlrrlr}
\hline
\hline
& \multicolumn{3}{c}{$0<z<2$} & \multicolumn{3}{c}{$2<z<6$} & \multicolumn{3}{c}{$0<z<6$}\\
& $\Delta z\pm \sigma_{\Delta z}$& \multicolumn{2}{c}{outliers} &$\Delta z\pm \sigma_{\Delta z}$  & \multicolumn{2}{c}{outliers} & $\Delta z\pm \sigma_{\Delta z}$ & \multicolumn{2}{c}{outliers}\\
\hline
HDF-N & $0.0\pm0.066$ & & & $-0.044\pm0.068$ & & & $-0.009\pm0.068$\\
& $0.0\pm0.066$ & 0/118 & & $-0.044\pm0.068$ & 1/33 & (3\%) & $-0.007\pm0.064$ & 1/151 & ($<$1\%)\\
\\
HDF-S & $0.033\pm0.378$ & & & $-0.017\pm0.075$ & & & $0.026\pm0.353$\\
& $0.033\pm0.063$ & 2/71 & (3\%) & $-0.017\pm0.075$ & 0/10 & & $0.030\pm0.065$ & 2/81 & (2\%)\\
\\
HDFs & $0.012\pm0.230$ & & & $-0.037\pm0.070$ & & & $0.002\pm0.210$\\
& $-0.011\pm0.066$ & 2/189 & (1\%) & $-0.037\pm0.070$ & 0/43 & & $-0.015\pm0.065$ & 3/232 & (1\%)\\
\hline
\end{tabular}
\end{table*}

\begin{figure}
\centering
\includegraphics[width=8cm,height=8cm]{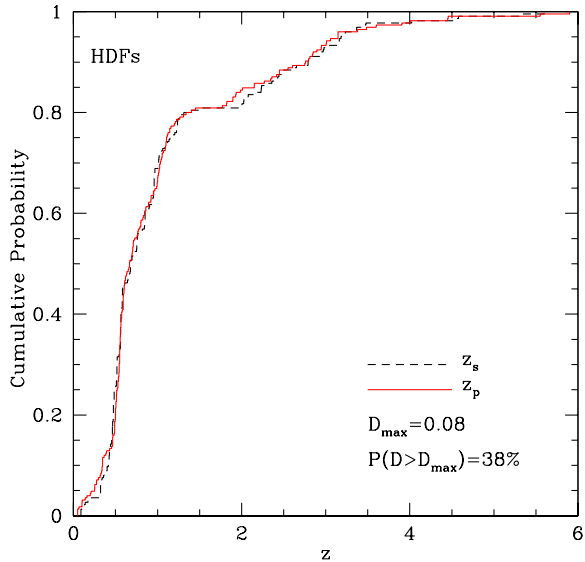}
\caption{Cumulative distributions of photometric (continuous line) and 
spectroscopic (dashed line) redshift of the 232 galaxies with known redshift.
The K-S test performed shows that the two distributions belong to
the same parent population.   
}
\end{figure}

\subsection{Redshift and color distributions} 
Once defined the set of templates and  parameters providing the most 
accurate estimate of the photometric redshift we have run the $\chi^2$
minimization procedure on the complete sample of 300 galaxies at  Ks$\le$23.
For those galaxies  having a spectroscopic redshift (74 at Ks$\le$23), the
best fitting has been constrained to the observed value.
\begin{figure}
\centering
\includegraphics[width=8cm,height=8cm]{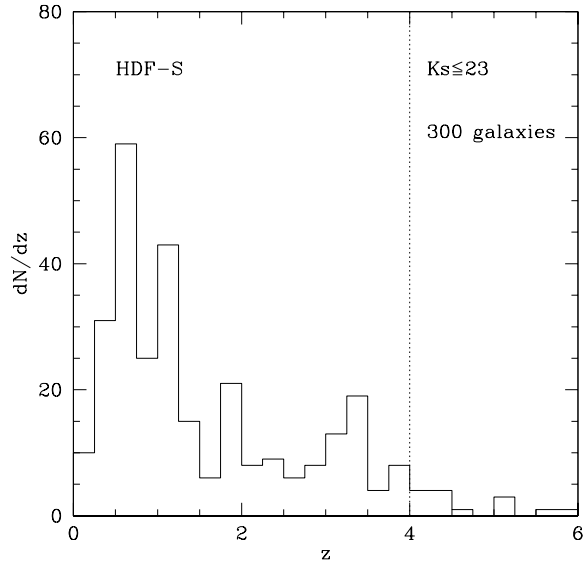}
\caption{Redshift distribution of the complete sample of 300 galaxies
brighter than Ks=23 in the HDF-S. 
The median redshift is $z_p=1.15$.
The dotted line at $z_p=4$  marks
the redshift limit of the highest redshift bin considered in the LF
estimate (see \S 5).}
\end{figure}
In Figure 7  the redshift distribution of the sample is shown.
The distribution has a median redshift $z_{m}=1.15$ and has a tail
extending up to $z_p\simeq6$.
There are 14 galaxies with photometric redshift $z_p>4$, 5 of which 
at $5<z_p\le6$. 
The Js-Ks color of the 300 galaxies is shown in Fig. 8.
Three of the Js-Ks$>$3 galaxies were noticed on shallower 
near-IR images (Saracco et al. 2001) and previously analysed 
resulting in high-mass evolved galaxies at $2<z<3$ (Saracco et al. 2004).
In Fig. 8 we also plot the  expected Js-Ks color obtained in the case
of E(B-V)=0 assuming a declining star formation rate with time scale 
$\tau=1$ Gyr seen at 5, 2 and 0.5 Gyr (from the top, thick lines) and  a 
constant star formation (cst)  seen at 1 Gyr (dot-dashed line).
The 1 Gyr old cst model is reddened by 
E(B-V)=0.5 to account for dusty star-forming galaxies (dot-dashed line).
Most of the  galaxies lie within the region limited by the 5 Gyr old 
model toward the red and by the 0.5 Gyr old model toward the blue.  
The reddest galaxies of the sample (at $2<z<3$) can be  described by  
old stellar populations, i.e. populations with  age comparable to the 
age of the Universe at the relevant redshift.
\begin{figure}
\centering
\includegraphics[width=8cm,height=8cm]{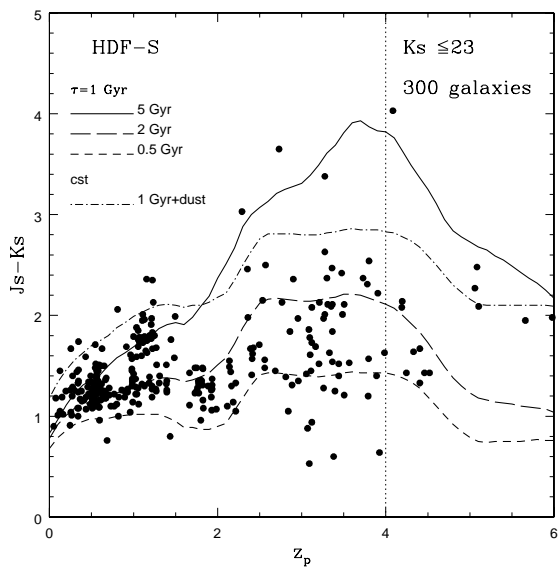}
\caption{Js-Ks color of galaxies versus redshift. 
The thick curves represent (from the top) the Js-Ks color derived by 
a declining star formation rate ($e^{-t/\tau}$) with e-folding time 
$\tau=1$ Gyr  and E(B-V)=0 seen at 5, 2 and 0.5 Gyr.
The thin curve is a dusty (A$_V$=2) constant star formation (cst) seen 
at 1 Gyr.
The templates have been obtained with a Salpeter IMF at solar metallicity.
The dotted line at $z=4$  marks the redshift limit of the highest 
redshift bin considered in the LF estimate (see \S 5).}
\end{figure}

\section{k-corrections and near-IR luminosities}
The estimate of the k-correction and of the absolute magnitude in the
near-IR rest-frame is a critical step when dealing with high redshift galaxies.
Indeed, unless having photometry sampling the near-IR rest-frame, 
it implies an extrapolation with respect to the observed wavelength.
This extrapolation is usually based on the best-fitting template which 
should be a good approximation of the true (but unknown) SED of 
the galaxy.
In fact, the larger the redshift, the wider the extrapolation and, 
possibly, the uncertainty affecting the rest-frame near-IR luminosity. 
Moreover, this estimate can also be affected by systematics 
due to the different libraries of models which can differ substantially in the 
near-IR domain.
We have tried to assess  whether and how  such 
uncertainties  affect the estimate of the rest-frame near-IR luminosity.
We first checked the robustness of the rest-frame near-IR luminosity 
with respect to the extrapolated photometry and than with respect to the different
models.
For each galaxy, we compared the rest-frame Js-band absolute 
magnitude obtained from the equation
\begin{equation}
M_{J_s}(k_J)=J_s - 5log(D_L(z))-25-k_J(z)
\end{equation}
where $D_L(z)$ is the luminosity distance at redshift $z$ in units of Mpc,
to the one derived by the equation
\begin{equation}
M_{J_s}(k_{JY})=m_Y - 5log(D_L(z))-25-k_{JY}(z)
\end{equation}
In eq. 1  $k_J(z)=(J_{s,rest}-J_s(z))_{temp}$ is the conventional k-correction
obtained as the difference between the rest-frame and observed magnitude 
computed on the best-fitting template.
To this end, the transmission curve of the ISAAC-Js filter  has been 
multiplied with the best fitting template at rest and redshifted to $z$
(where $z=z_s$ for the 74  galaxies with known redshift in the K23 sample).
In eq. 2, $k_{JY}=(J_{s,rest}-Y(z))_{temp}$ and $m_Y$  is the observed 
apparent magnitude in the filter Y that best matches the rest-frame 
Js-band of the galaxy at the relevant redshift  (e.g. Lilly et al. 1995; 
Pozzetti et al. 2003).
In our case  Y=Js, H, Ks according to the redshift of the galaxy.
Both the ``color k-correction'' terms,$k_J(z)$ and $k_{JY}$,  
include the flux density dimming 
factor $(1+z)$  independent of wavelength.
\begin{figure*}
\centering
\includegraphics[width=8.5cm,height=8.5cm]{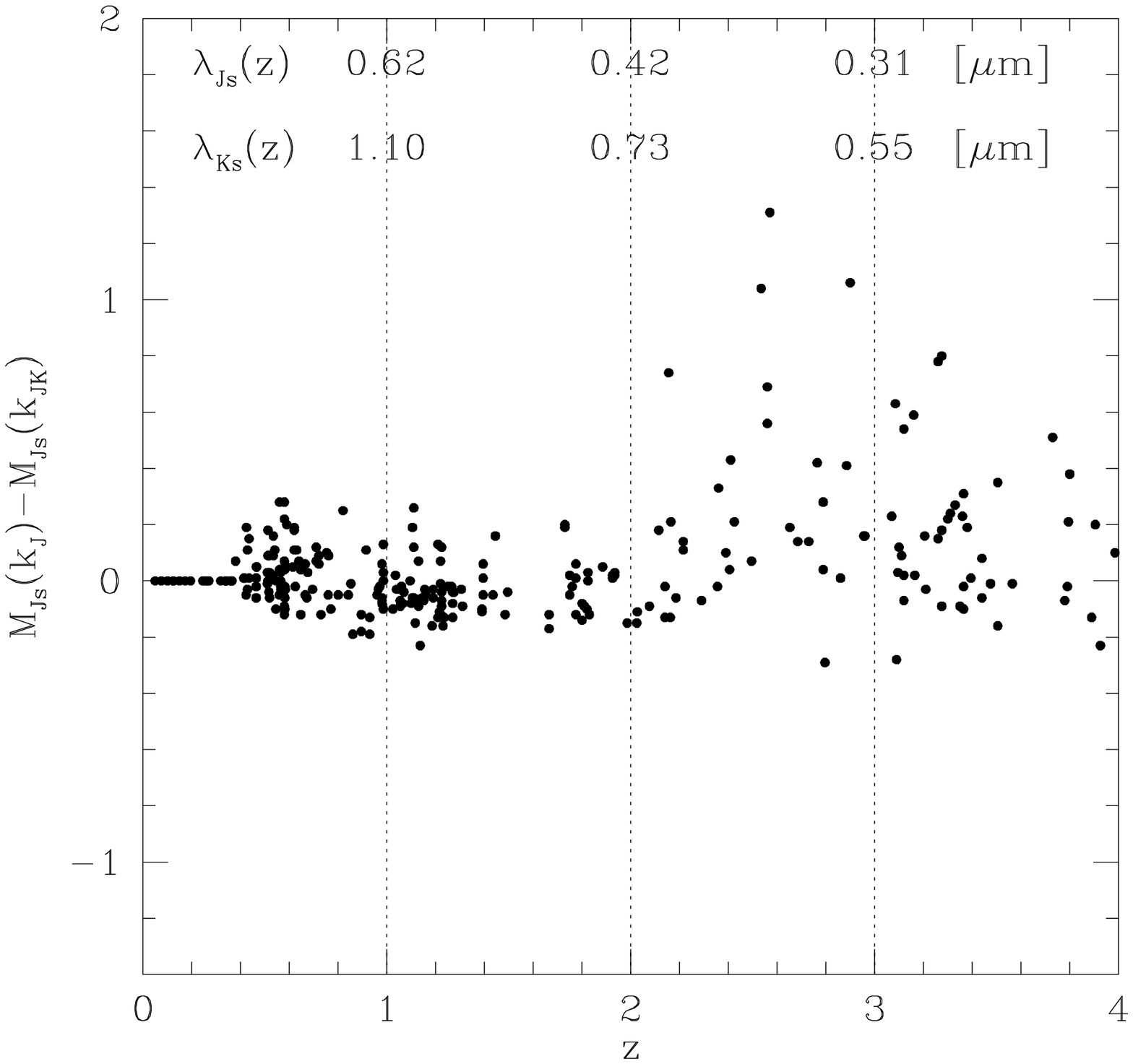}
\includegraphics[width=8cm,height=8cm]{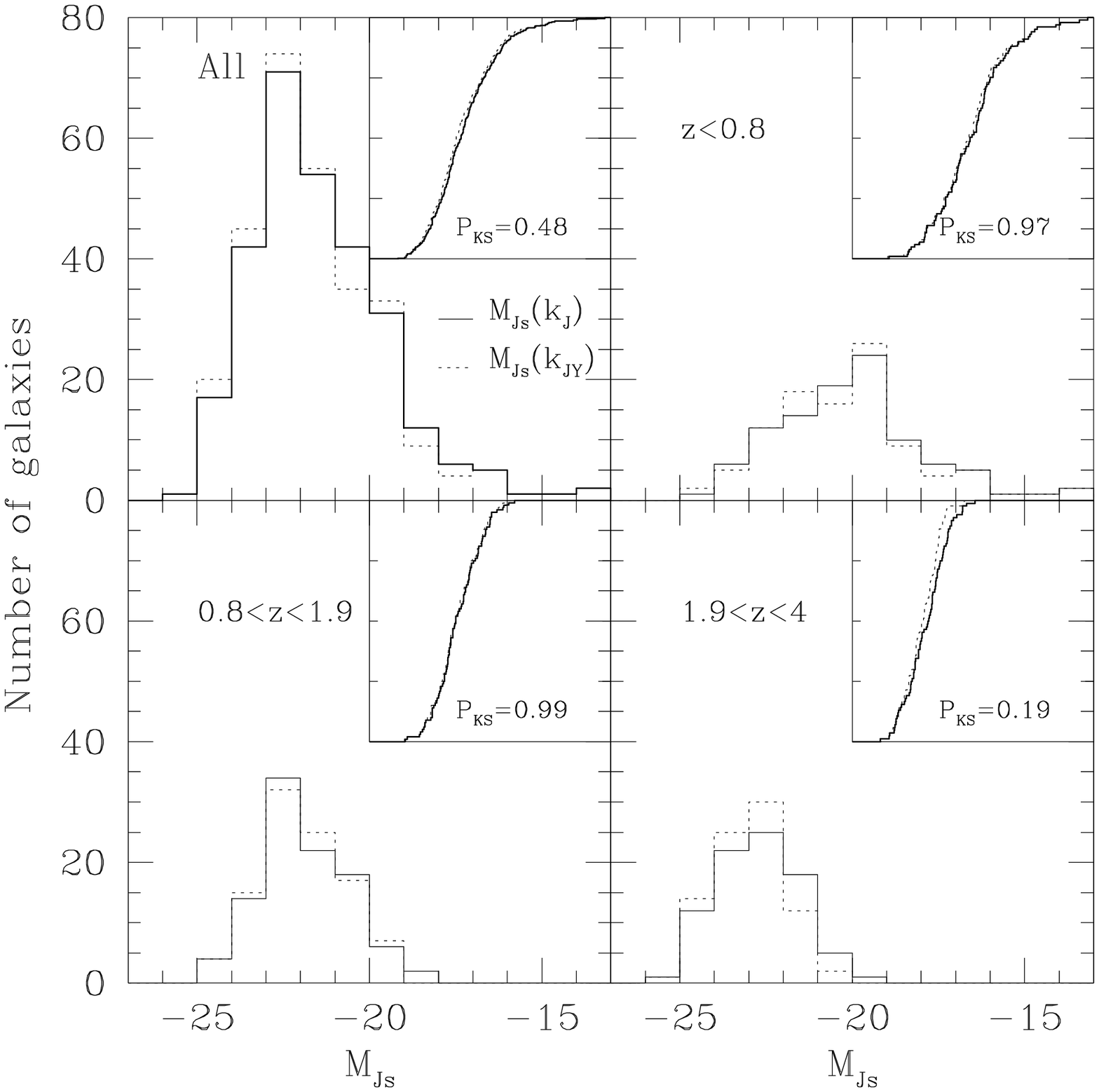}
\caption{{\em Left}: The difference between the
rest-frame Js-band absolute magnitude derived for each galaxy from eqs.1 
and 2 is plotted as a function of $z$. 
{\em Right}: The distribution of the Js-band absolute magnitude of galaxies 
derived from eq. 1 (solid histogram) are compared 
to the one derived from eq. 2 (dotted histogram) for different redshift bin. 
No significant differences are present between the distributions at $z<2$
as confirmed by the high probability provided by the K-S test 
(small sub-panels). }
\end{figure*}
In Fig. 9 (left panel) the difference between the rest-frame Js-band absolute 
magnitude derived for each galaxy from equations 1 and 2 is plotted as 
a function of $z$.
Up to $z\sim2$, where  the near-IR filters  sample
wavelengths $\lambda_{Js}>0.4$ $\mu$m and $\lambda_{Ks}>0.7$ $\mu$m, 
the two equations 
provide the same values within an rms of about $0.15$ mag, 
comparable to or lower than the typical photometric error.
This dispersion does not affect the estimate of the LF.
This is shown in the right panel of Fig. 9 where the distributions of the 
rest-frame Js-band absolute magnitude derived from eq. 1 (solid histogram) 
and from eq. 2 (dotted histogram) are compared for galaxies in different 
redshift bin. 
At $z<2$, the two distributions belong to the same parent population as 
confirmed by the K-S test we performed.
At larger redshift, the conventional eq. 1 
provides absolute  magnitudes which can differ  
systematically from those derived from eq. 2. 
These differences can bias the estimate of the LF.
The absolute magnitude distributions shown in Fig. 9 for galaxies 
at $z>2$ are indeed rather different, even if the difference is only 
marginally significant.
Thus, until the near-IR band chosen to construct the LF (the Js filter in 
our example) samples the part of the SED red-wards $\lambda>0.4$ $\mu$m, 
characterized by a regular shape, dominated by the emission of older 
stars and weakly affected by dust extinction and star formation, 
eq. 1 and 2 provide consistent absolute magnitudes.
In this case, having photometry approaching the rest-frame wavelength of the
chosen filter (the Ks-band in our example) and using eq. 2 do not improve 
the accuracy and the reliability of the estimate of the LF.
On the contrary, when the chosen near-IR filter samples the blue and UV part 
of the spectrum ($\lambda_{Js}<0.4$ $\mu$m at $z>2$), the observed emission 
in that filter (Js) is dominated by young ($<0.5$ Gyr) stars which do not 
contribute significantly to  the red part of the spectrum, i.e. to the
rest-frame luminosity we want to derive. 
Moreover, it can be strongly affected by dust extinction.
In this case the extrapolation to the near-IR rest-frame wavelength  
required by eq. 1 can give wrong values and the photometry in the 
redder band, the Ks filter in our case,  is needed to apply eq. 2.

We then checked the dependence of the LF on the different library of
models used to derive the best-fitting template.
Recently, Maraston (2005; M05 hereafter) has shown the importance of the 
TP-AGB phase  in modeling young stellar populations. 
This evolutionary phase can play an important role in the continuum 
emission of a stellar population at $\lambda>0.4-0.5$ $\mu$m  with ages 
younger than 1 Gyr.
The modeling of this phase in the Maraston's code produces a brightening 
in the near-IR with respect to the same stellar populations
(equal age, IMF and metallicity) modeled by the Bruzual \& Charlot (2003)  
and  PEGASE (Fioc \& Rocca-Volmerange 1997) codes.
To assess  whether  this difference affect the estimate of 
the LF, we compared the distributions of the Ks-band absolute magnitudes  
obtained with the BC03 models and  with the M05 models.
We did not include the PEGASE models in this comparison since they provide
spectral energy distributions very similar to those of BC03.
We constructed the same set of templates described in Tab. 1 using
the SSPs of M05 and we searched for, for each galaxy, the best-fitting 
template at the relevant redshift. 
In Fig. 10 the  absolute magnitude distributions obtained 
with the two sets of models for the whole sample of galaxies at
$z<4$ and in three redshift bin are shown.
The K-S test we performed does not show evidence of  
difference between the distributions providing
 probabilities  P($_{KS}$)$>0.8$ in all the redshift bin.

The above results show that our study of the LF and of its evolution 
with redshift is not dependent either on the method used to derive the 
rest-frame absolute magnitudes or on the library of templates used at 
least down to $z\simeq2$.
Therefore, we are confident of the reliability of our estimate of the LF at 
least down to this redshift.
However, we  chose to push our study of the LF at $z<4$ since the Ks-band 
still samples the red part ($\lambda_{Ks}>0.45$ $\mu$m) of the SED down 
to this redshift limit.
We will use eq. 2 to derive the rest-frame Js-band absolute magnitudes
and the conventional form of eq. 1 to compute  the Ks-band absolute magnitudes.
In Fig. 11 the Ks-band k-correction derived for the sample of 285 galaxies at 
$z<4$ is shown as a function of redshift. 
For comparison, superimposed on the data, we plot the k-correction as 
derived by 3 synthetic templates (continuous lines) obtained with a declining
star formation rate seen at three different ages, (from the top) 
5, 2 and 0.5 Gyr. 
Besides the models, we also plot (dashed line) the k-correction derived 
by the mean observed spectrum of local elliptical galaxies 
(Mannucci et al. 2001).
\begin{figure}
\centering
\includegraphics[width=8cm,height=8cm]{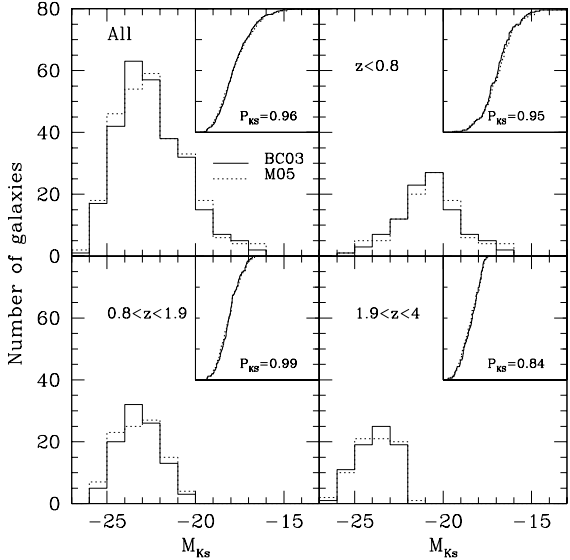}
\caption{The K-band absolute magnitudes obtained with the
BC03 (Bruzual \& Charlot 2003) models (solid histogram) for the sample of 
285 galaxies at $z<4$ are compared with those obtained with the M05 
(Maraston 2005) 
models (dotted histogram) for different redshift bins. 
The distributions belong to the same parent population as shown by the 
high probability provided by the K-S test we performed (small panels).
}
\end{figure}
\begin{figure}
\centering
\includegraphics[width=8cm,height=8cm]{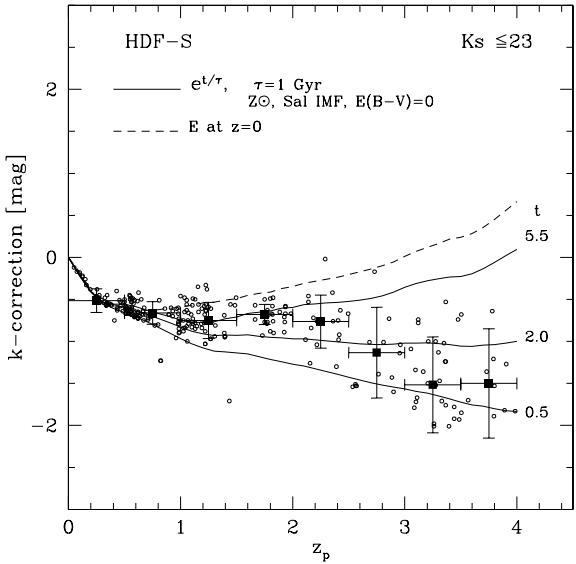}
\caption{Ks-band k-correction as a function of redshift for  the 285 galaxies
brighter than Ks=23 at $z\le4$ in the HDF-S (small open circles). 
The filled squares are the median values of the k-correction in redshift 
bins 0.5 width (horizontal errorbars).
The errorbars along the y-axis represent the scatter within each bin.
The continuous lines are the k-corrections derived by a model described by
an exponentially declining star formation rate  with
$\tau=1 Gyr$ and t (from top) 5, 2 and 0.5 Gyr.
The dashed line represent the k-correction of local ellipticals  derived
by the mean observed spectrum of ellipticals (Mannucci et al. 2001). 
}
\end{figure}

\section{The near-IR Luminosity Function in the HDF-S}
\begin{figure}
\centering
\includegraphics[width=9cm,height=11cm]{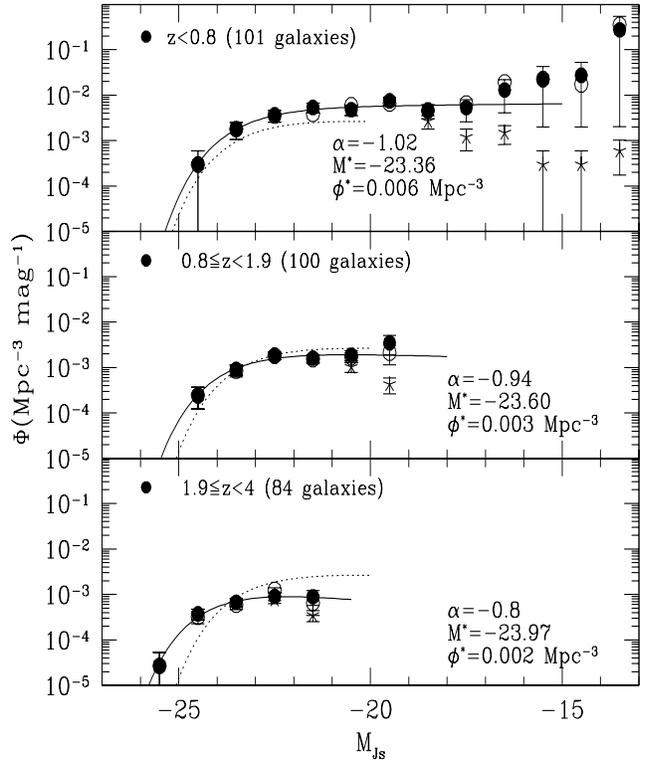}
\caption{Rest-frame Js-band luminosity function of galaxies derived with the 
1/$V_{max}$ method (circles) in the three redshift bin considered: 
$0<z<0.8$ ($z_{m}=0.55$, upper panel),  $0.8\le z<1.9$ ($z_{m}=1.2$, middle panel) and $1.9\le z<4$ ($z_{m}=3.1$, lower panel).
We marked with open circles the LF obtained with the Js-band luminosities 
derived from eq. 1. The starred symbols show the LF uncorrected for the
V$_{max}$.
The faint end (M$_{Js}>-17$) of the LF is defined by 9 galaxies at $z<0.3$.
Superimposed on the LF is shown the formal fit obtained with a Schechter 
function (thin line). The dotted line is the local LF of Cole et al. (2001).}. 
\end{figure}
\begin{figure}
\centering
\includegraphics[width=9cm,height=11cm]{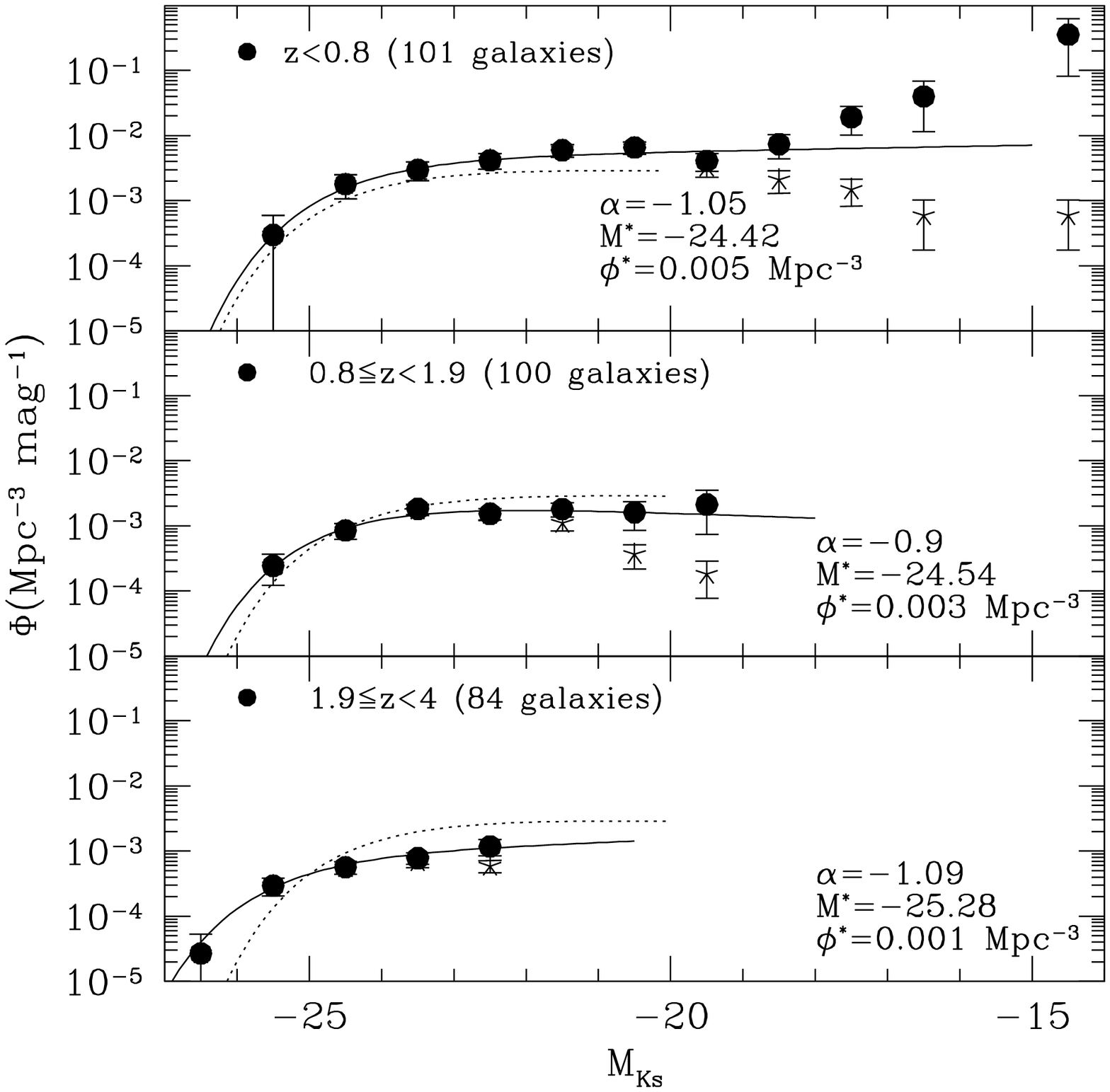}
\caption{Rest-frame Ks-band luminosity function of galaxies  and
Schechter best-fitting. Symbols are as in Fig. 12.}
\end{figure}

\subsection{Computing the LF with $1/V_{max}$ method}
We computed the luminosity function, $\Phi(M)dM$, using the non-parametric
$1/V_{max}$ method (Schmidt 1968).
This method does not require any assumption on the form of the LF and
provides an unbiased estimate  if the sample is not affected by strong density
inhomogeneities.
Previous studies of the LF of galaxies in the HDFs 
obtained with different estimators, both parametric and non-parametric 
(Takeuchi et al. 2000; Bolzonella et al. 2002),  show  no 
differences with respect to the LF obtained with the $V_{max}$ formalism.
This implies that no strong inhomogeneities are present in the HDF-S and that
 our results are not dependent on the LF estimator used.
In the $1/V_{max}$ method, given a redshift bin $[z_l,z_u)$, each galaxy contributes 
to the number density of galaxies  in that bin an amount inversely proportional to the 
maximum volume:
\begin{equation}
V_{max,i}=\int_\Omega \int_{z_l}^{min(z_u,z_{max,i})} {d^2V\over dz d\Omega}dz d\Omega 
\end{equation}
where $z_{max,i}$ is the maximum redshift at which the galaxy $i$ with 
absolute magnitude $M_{i}\in[M, M+dM]$ is still 
detectable given the limiting apparent magnitude Ks=23  
of the sample, ${d^2V/dz d\Omega}$ is the comoving volume element and 
$\Omega$ is the solid angle covered by the surveyed area.
In our case, the sample has been selected over an area of about 5.5 arcmin$^2$
of the HDF-S corresponding to a solid angle 
$\Omega\simeq4.65\times10^{-7}$ std.
The comoving number density of galaxies $\Phi(M)dM$ for each absolute magnitude
bin in a given redshift bin and its error $\sigma_{\Phi}$ are computed as
\begin{equation}
 \Phi(M)dM= \sum_i{1\over{V_{max,i}}}dM ~~~
\sigma_{\Phi}=\sqrt{\sum_i\left({1\over{V_{max,i}}}\right)^2}
\end{equation}
We defined the  different redshift bins so that  a comparable and 
statistically significant number of galaxies fall in each of them.
According to this criterion and to the results obtained in \S 4,
we divided the sample of 285 galaxies  at $z<4$ in the three redshift bins 
[0;0.8), [0.8;1.9) and [1.9;4). 
Each bin includes  101, 100 and 84 galaxies   of which
52 (50\%), 12 (12\%) and 10 (12\%) with spectroscopic redshift respectively.
The median redshifts in the three bin are $z_{m}=0.55$, $z_{m}=1.2$ and
$z_{m}=3.1$ respectively.

The rest-frame Js-band and Ks-band LFs we derived in each redshift bin 
are shown in Fig. 12 and 13 respectively (filled points).
The starred symbols denote the LF uncorrected for the incompleteness
and show where the 1/V$_{max}$ correction takes place.
In Fig. 12, the Js-band LF obtained with the rest-frame Js-band absolute 
magnitudes derived from eq. 1 is also shown (empty points).
The figures suggest a systematic decrease of the number density 
of bright  galaxies coupled with a 
systematic  brightening of the LF going to high redshift.
This trend is present both in the Js-band LF and in the Ks-band LF. 
We will probe further these features in the next section.
{ Besides this, the figures are also suggestive of a raise of the LF}
at faint absolute magnitudes (M$_{Js}>-17$ and M$_{Ks}>-18$) 
in the lowest redshift bin, raise which cannot be probed in the two bins 
at higher redshift where  
brighter galaxies are selectively missed.
{ The same slope of the faint end is obtained by computing the LF in the 
redshift range $0<z<0.4$ suggesting that the raise is not due to possible  
exceeding values of 1/V$_{max}$. 
However, it is worth noting that the raise we see
is based on a very low statistics.
Indeed, the faint end is defined by 9 out of the 12 galaxies at $0.1<z<0.3$
in our sample.} 
It is indeed expected that, given the criteria adopted to select the HDF-S 
(devoid of bright sources),  galaxies at low redshift 
in this field are low luminosity galaxies (M$_K>$-20).
This, coupled with the very faint limit (Ks=23) of the sample, allows to
probe the faint end of the luminosity function at 
$\langle z\rangle\simeq0.2$ down to the unprecedented faint near-IR magnitude
M$\simeq-14$.
These low luminosity galaxies are  slightly bluer in the optical 
(0.45$<$V$_{606}$-I$_{814}<0.9$) than the bulk of the galaxies in the HDF-S 
while their optical-IR colors (1.4$<$I$_{814}$-Ks$<2.7$, 0.9$<$Js-Ks$<$1.5) 
are consistent with those of the bulk.
All but one  are brighter than V=25.8 and I=24.9 
and are in the range  21$<$Ks$<22.9$. 
Most of them  show an irregular morphology while two of them appear very 
compact.
In  Fig. 14  the V$_{606}$ (upper panels) and the 
I$_{814}$-band  (lower panels) images of 7 out of the 9 galaxies at 
$z<0.3$ fainter than M$_K$=-18 are shown.
Both the { apparent} raise of the LF at faint luminosities and the 
irregular morphology 
of the galaxies populating the faint end are consistent with  previous 
estimates of the local LF of galaxies at optical wavelengths.
Such estimates show that the faint end of the near-IR LF is
dominated by irregular and dwarf galaxies whose contribution to 
the comoving number density increases going to lower luminosities 
as found for the local LF in the optical rest-frame
(e.g. Marzke et al. 1994;  Marzke et al. 1998; Zucca et al. 1997; 
Folkes et al. 1999).  
We estimated a comoving number density of galaxies with magnitude 
-19$<$M$_K<$-14 of $\bar n_{faint}=0.4\pm0.27$ Mpc$^{-3}$ at 
$\langle z\rangle\simeq0.2$. 
This value agrees with the one derived locally by Zucca et al. (1997) for galaxies
M$_{b_j}>-17$ and it is nearly one order of magnitude higher
than the comoving density of brighter galaxies 
($\bar n_{bright}=0.039\pm0.005$ Mpc$^{-3}$ for M$_K<-19$).
{ However, the very low statistic, the uncertainty in the photometric 
redshift estimate and the small volume sampled
by the HDF-S, especially at low redshift,  prevent us to strongly constrain 
the faint end of the LF. 
The possible  raise we see at 
very faint near-IR luminosities  needs a larger sample to be established.}
\begin{figure*}
\centering
\includegraphics{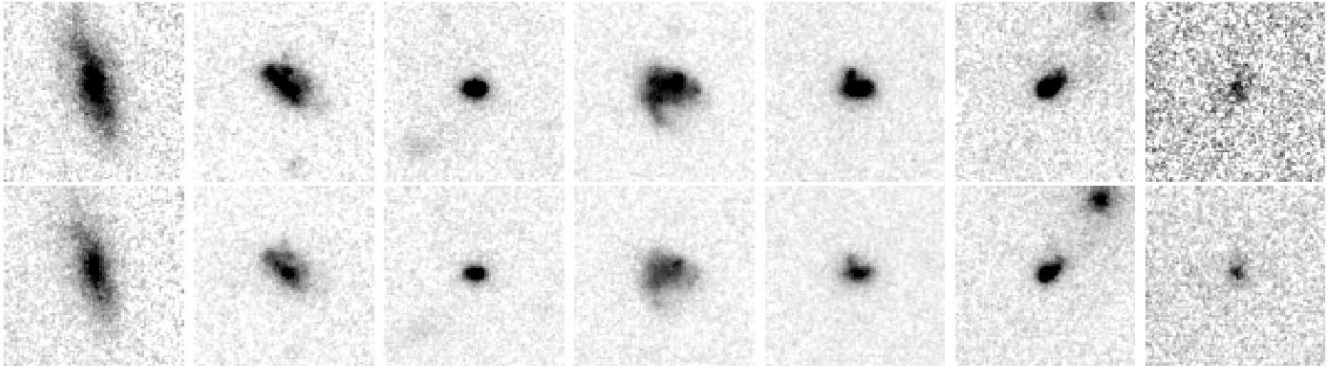}
\caption{V$_{606}$-band (upper panels) and I$_{814}$-band (lower panel) 
images of 7 out of the 9 galaxies at $z<0.3$ fainter than M$_K$=-19 galaxies.
The images are 3$\times$3 arcsec centered on the source.
The intensity of the images are optimized to show the galaxies.
Galaxies are displayed from left to right on the basis of their 
V$_{606}$-band apparent magnitude.}
\end{figure*}

\subsection{The evolution of the LF of galaxies in the HDF-S} 
In Figs. 12 and 13 we  plot the formal fitting to the LFs we obtained with a 
Schechter function (Schechter 1976).
The fitting  is performed over the whole range of absolute magnitude
spanned by the LFs in each redshift bin.
The results of the fitting are summarized in Tab. 3 where we report the
values of the  parameters $\alpha$, M$^*$ and $\phi^*$ in each redshift
bin.
\begin{table}
\centering
\caption{
Parameters  $\alpha$, M$^*$ and $\phi^*$ of the Schechter function 
obtained by fitting the Js-band and the Ks-band LFs in the three redshift bin.
}
\begin{tabular}{l c c c}
\hline
\hline
$z$-bin & $\alpha$ & M$^*$ & $\phi^*$ \\
  &          &       & [10$^{-3}$Mpc$^{-3}$]\\  
\hline
	&	& Js-band & \\
\hline
0.0-0.8 & $-1.02^{+0.10}_{-0.10}$ & $-23.36^{+0.63}_{-0.57}$ & $6.0^{+2.3}_{-1.6}$ \\
0.8-1.9 & $-0.94^{+0.16}_{-0.15}$ & $-23.60^{+0.35}_{-0.36}$ & $2.6^{+1.1}_{-1.0}$ \\
1.9-4.0 & $-0.81^{+0.34}_{-0.25}$ & $-23.97^{+0.50}_{-0.52}$ & $2.0^{+0.8}_{-0.6}$ \\
\hline
	&	& Ks-band & \\
\hline
0.0-0.8 & $-1.05^{+0.11}_{-0.10}$ & $-24.42^{+0.65}_{-0.58}$ & $5.2^{+2.4}_{-1.5}$ \\
0.8-1.9 & $-0.90^{+0.18}_{-0.15}$ & $-24.54^{+0.35}_{-0.38}$ & $2.8^{+1.1}_{-0.9}$ \\
1.9-4.0 & $-1.09^{+0.34}_{-0.27}$ & $-25.28^{+0.52}_{-0.54}$ & $1.0^{+0.7}_{-0.5}$ \\
\hline
\end{tabular}
\end{table}
In Fig. 15 the errors contour at 68\% confidence level for the joint 
parameters [$\alpha$, M$^*$] and  [$\phi^*$, M$^*$] of the Ks-band LF fit
are shown.
\begin{figure*}
\centering
\includegraphics[width=7cm,height=6.5cm]{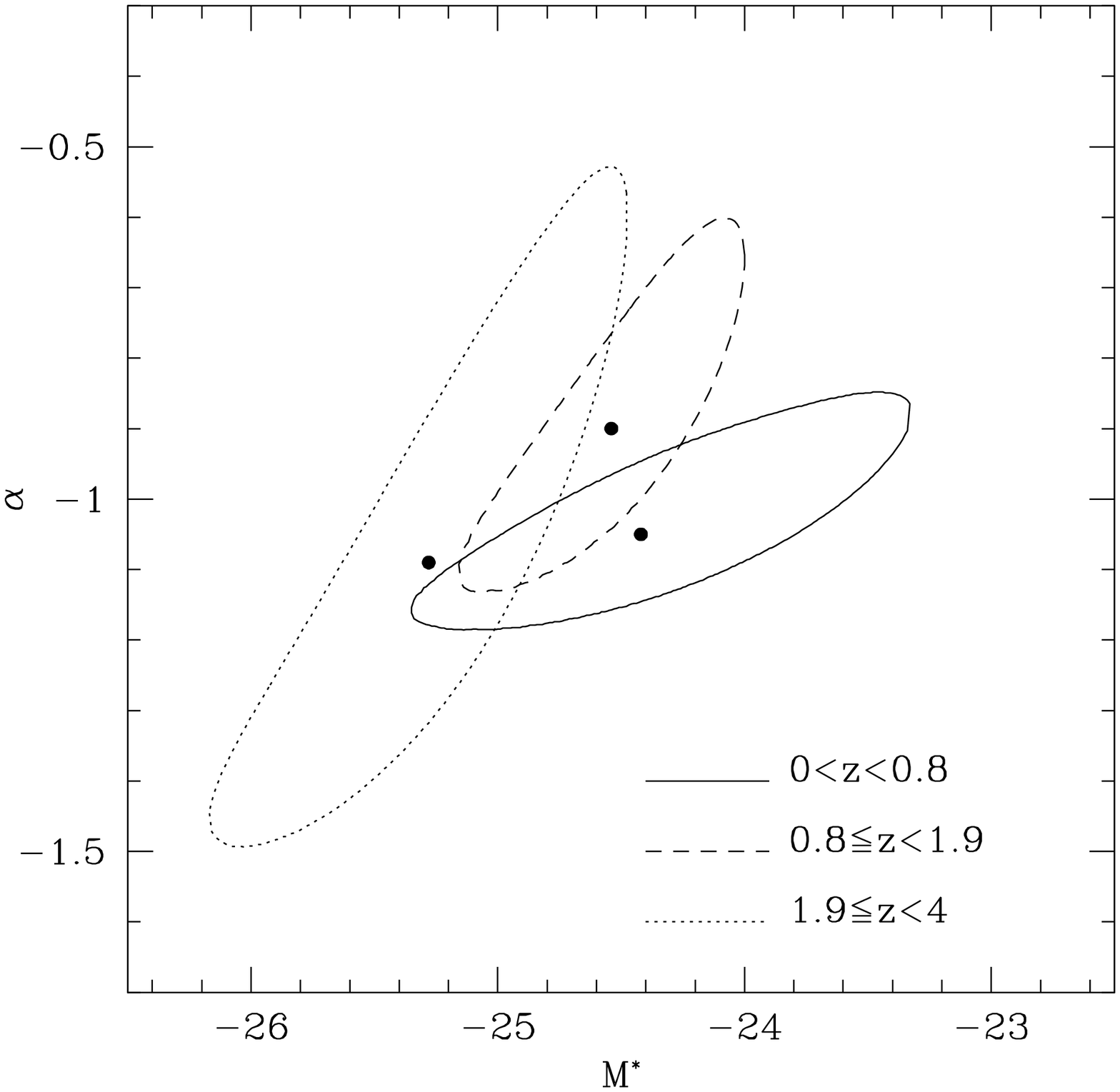}
\includegraphics[width=7.4cm,height=6.5cm]{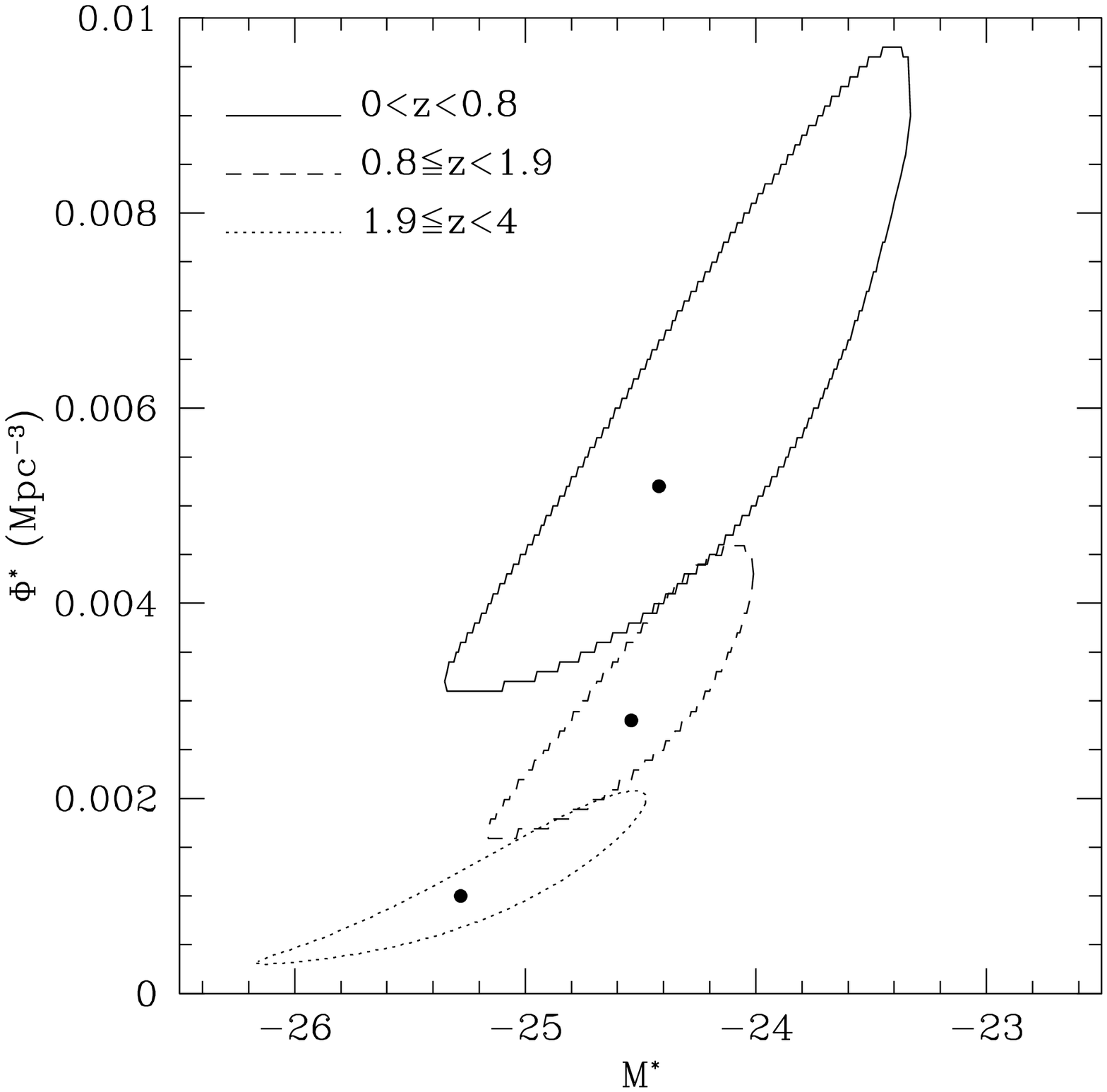}
\caption{Error contours at 68\% confidence level 
for the  parameters [$\alpha$, M$^*$] (left panel) and [$\phi^*$, M$^*$]
(right panel) of the Ks-band Schechter fit of Fig. 13.}
\end{figure*}
{ The decreasing number density and the possible brightening previously 
noticed in the LFs going to high redshift  are evident in the plot.
However, given the statistical correlation between the Schechter parameters 
($\alpha$, M$^*$ and $\phi^*$), we will probe further on the significance of 
this evolution by fixing the $\alpha$ parameter.}
Fig. 15 suggests also that we cannot constrain the faint 
end of the LF and thus the parameter $\alpha$ at high redshift.
However,  we  tried to constrain the evolution of 
the LF  shape  by means of the K-S test (see e.g. Feulner et al. 2003)
which uses all the absolute magnitudes without binning. 
We derived in each redshift bin the cumulative LF (CLF)
which gives the number of galaxies per unit volume brighter than M.
In the upper panel of Fig. 16 we show the three CLFs obtained in the Ks-band.
Since the  K-S test compares the CLFs normalized to unit, this test is
not able to detect  variations of $\phi^*$.
The normalized CLFs provide information on the relevant distribution
of galaxies within the range of absolute magnitude considered, 
i.e. on the form of the LF.
For this reason we expect that the K-S test is more sensitive to 
the variation of $\alpha$ which defines the form  of the distribution 
rather than of M$^*$.

To quantify the sensitivity of the K-S test to the variations
of  $\alpha$ and M$^*$ for samples of the same size of our sample we simulated
and compared among them  various samples of 100 objects described by  a  
Schechter LF with different values of $\alpha$ and M$^*$. 
We have found that, at  90\% confidence level, the minimum 
variations we are able to detect on $\alpha$ and M$^*$ on our
samples are $\pm0.3$  and $\pm0.8$ mag respectively. 
In the lower panel of Fig. 16 the normalized CLFs obtained in the three 
redshift bin  are compared two by two (lowest vs mid redshift bin (upper panel)
 and  mid vs highest redshift bin (lower panel)).
\begin{figure*}
\centering
\includegraphics[width=7.4cm,height=7cm]{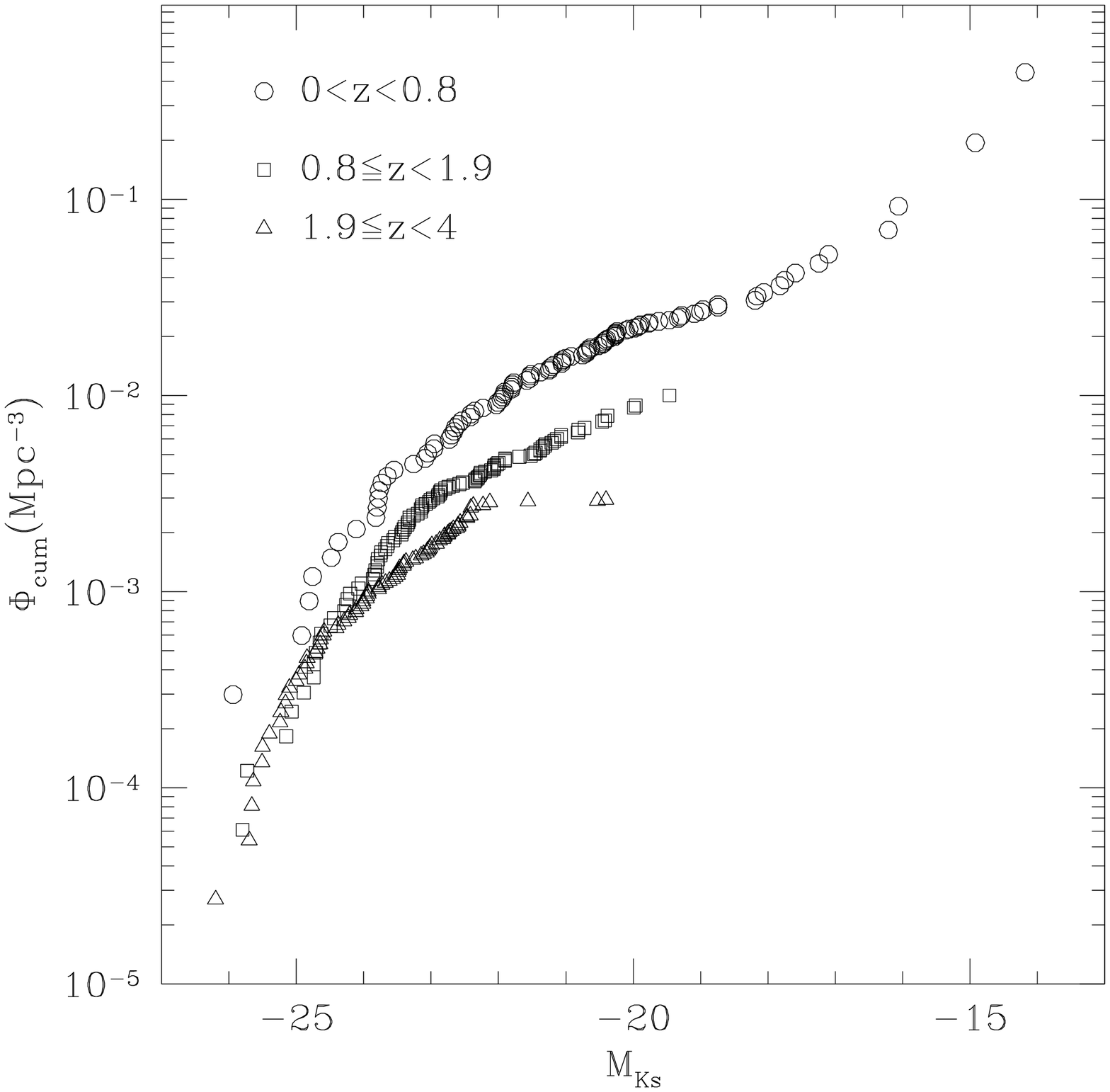}
\includegraphics[width=7.4cm,height=7cm]{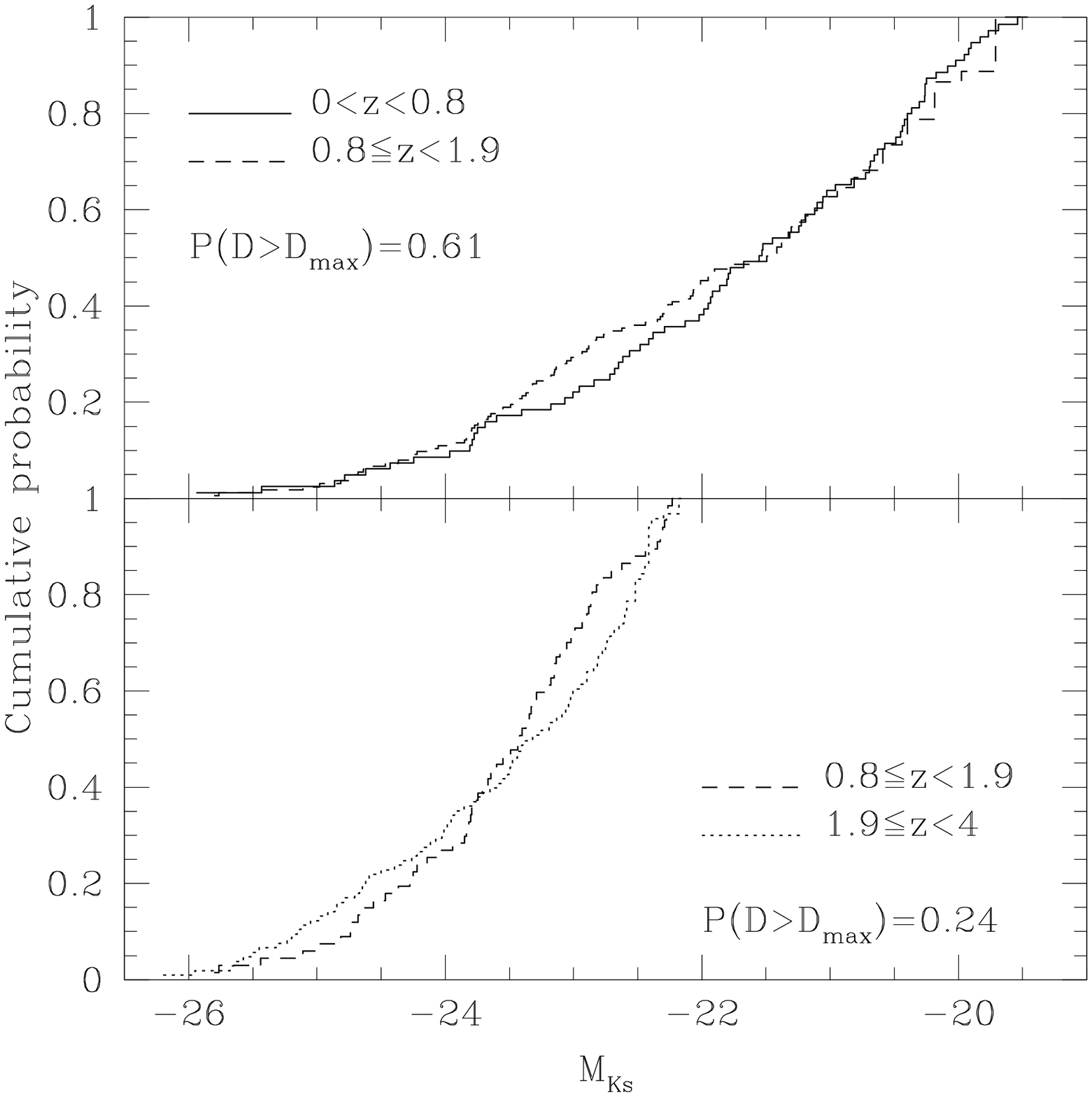}
\caption{{\em Left} - Cumulative luminosity function of galaxies derived 
with the 1/$V_{max}$ in the three redshift bin: $0<z<0.8$ (circles),  
$0.8\le z<1.9$ (squares) and  $1.9\le z<4$ (triangles).
{\em Right} - Two by two comparison of the cumulative LFs in the three 
redshift bin normalized over the range of absolute magnitude in common.}
\end{figure*}
The K-S test does not point out significant differences among the 
distributions as shown by the probabilities P$_{KS}\simeq$0.61 
(bin 1 vs bin 2) and P$_{KS}\simeq$0.24 (bin 2 vs bin 3).
Thus, we conclude that if $\alpha$ changes its variation has to be lower than 
0.3 from  $z_m\simeq0.6$ to $z_m\simeq3$.

We then  probed the evolution of M$^*$ and $\phi^*$ by comparing the values 
obtained by fitting a Schechter function for a fixed value of $\alpha$.
We assumed the value $\alpha=-1.0$ since it is very close to the values
we derived in each redshift bin and to the values previously
derived by other authors for local galaxies (e.g. Kochanek et al. 2001;
Loveday 2000).
\begin{figure*}
\centering
\includegraphics[width=8.cm,height=7.5cm]{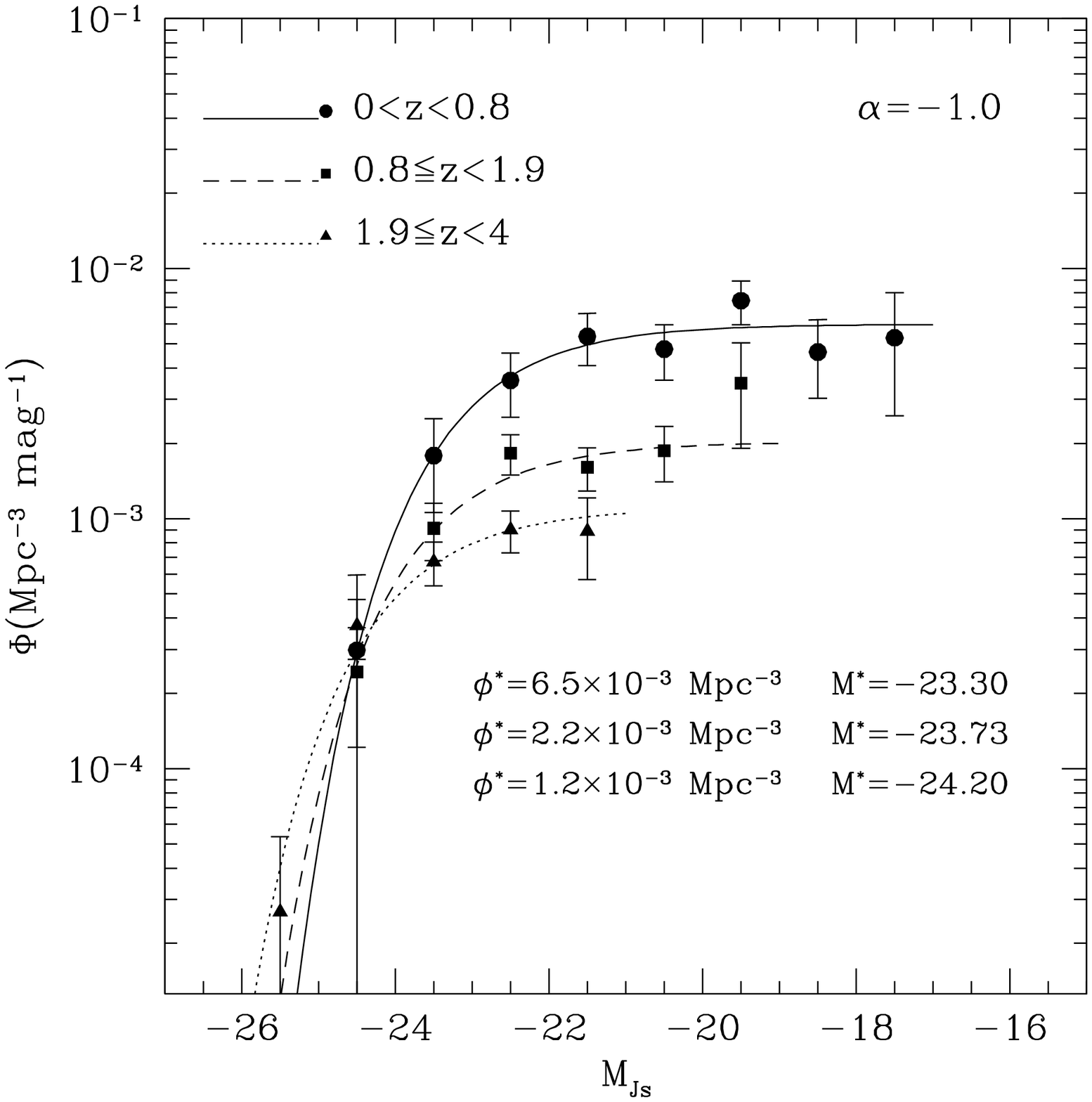}
\includegraphics[width=8.cm,height=7.5cm]{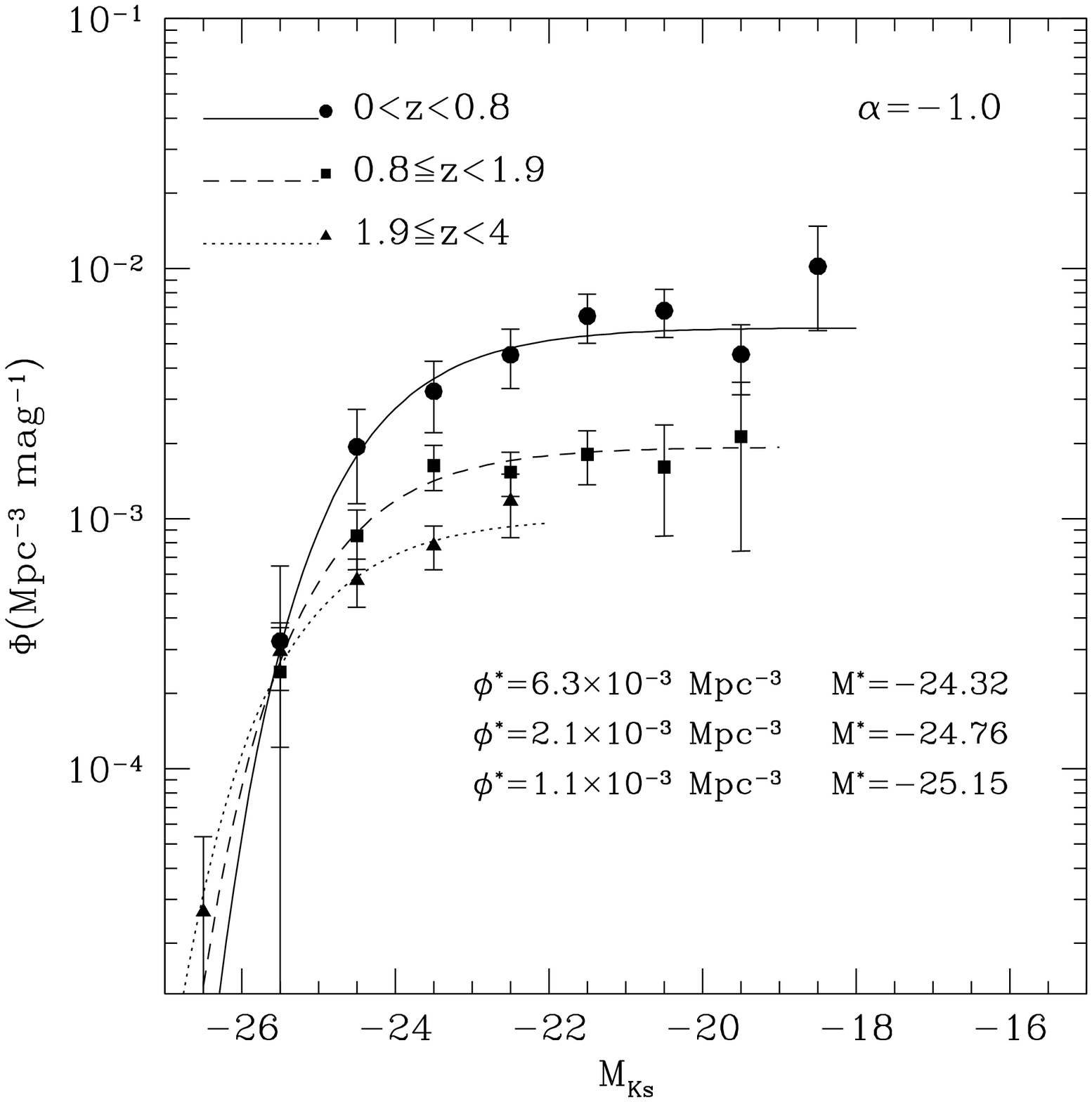}
\caption{{\em Left} -  Superimposed on the Js-band LFs are shown the relevant 
best fitting obtained with a Schechter function with $\alpha=-1.0$. 
{\em Right} - Same as in the left panel but for the Ks-band.}
\end{figure*}
\begin{table}
\centering
\caption{Parameters  M$^*$ and $\phi^*$  of the Schechter function 
fitting the Js-band and the Ks-band LF with $\alpha=-1.0$  and
 the Ks-band LF with $\alpha=-1.1$.
to the LF in the three redshift bin.}
\begin{tabular}{l c c c}
\hline
\hline
$z$-bin & $\Delta$M & M$^*$ & $\phi^*$ \\
  &          &       & [10$^{-3}$Mpc$^{-3}$]\\  
\hline
        &  Js      & $\alpha=-1.0$                 &   \\
\hline
0.0-0.8 & -25;-17& $-23.30^{+0.40}_{-0.40}$ & $6.5^{+1.0}_{-0.9}$ \\
0.8-1.9 & -25;-19& $-23.72^{+0.22}_{-0.22}$ & $2.2^{+0.2}_{-0.2}$ \\
1.9-4.0 & -26;-21& $-24.20^{+0.21}_{-0.20}$ & $1.2^{+0.2}_{-0.2}$ \\
\hline
        &  Ks      & $\alpha=-1.0$                 &   \\
\hline
0.0-0.8 & -26;-18& $-24.32^{+0.41}_{-0.40}$ & $6.3^{+1.1}_{-1.0}$ \\
0.8-1.9 & -26;-19& $-24.76^{+0.23}_{-0.23}$ & $2.1^{+0.2}_{-0.3}$ \\
1.9-4.0 & -27;-22& $-25.15^{+0.25}_{-0.23}$ & $1.1^{+0.2}_{-0.2}$ \\
\hline
        &   Ks     & $\alpha=-1.1$                 &   \\
\hline
0.0-0.8 & -26;-19& $-24.62^{+0.40}_{-0.39}$ & $4.5^{+1.1}_{-1.0}$ \\
0.8-1.9 & -26;-19& $-25.03^{+0.22}_{-0.21}$ & $1.6^{+0.2}_{-0.3}$ \\
1.9-4.0 & -27;-22& $-25.28^{+0.23}_{-0.20}$ & $0.9^{+0.2}_{-0.2}$ \\
\hline
\end{tabular}
\end{table}
\begin{figure}
\centering
\includegraphics[width=8.5cm,height=7.5cm]{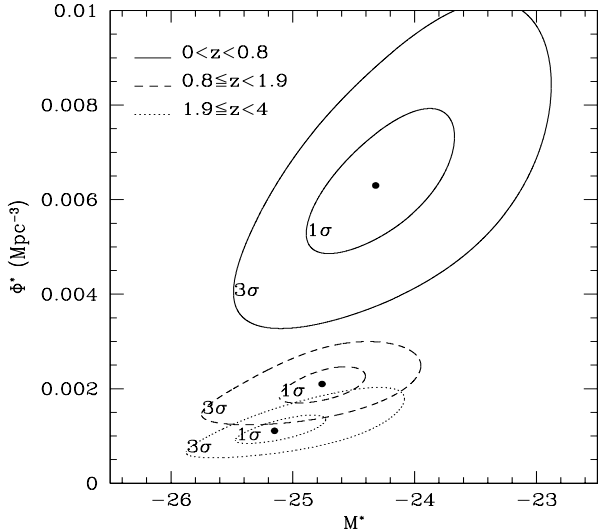}
\caption{Error contours at 1$\sigma$ and 3$\sigma$ for the 
parameters $\phi^*$ and M$^*$ of the Schechter fit with $\alpha=-1.0$ 
to the Ks-band LF.}
\end{figure}
The results of the fitting are summarized in Tab. 4 and shown in Fig. 17.
The possible brightening combined to the decreasing number density of 
bright galaxies previously noticed in the LF from $z_{m}\simeq0.6$ 
to $z_m\simeq1.2$ is confirmed and, possibly, it extends to $z_m\sim3$.
The normalization of the LF and, consequently, the number 
density of bright  galaxies decreases by a factor  3 from 
$z_m\simeq0.6$ to $z_m\simeq1.2$.
In parallel, the characteristic magnitude M$^*$ brightens by $\sim0.4$ mag. 
{ The evolution of $\phi^*$ is detected at  high confidence level 
($>3\sigma$), as shown by the error contours shown in Fig. 18 while 
the significance of the brightening of M$^*$ is $1\sigma$. }
The same trend persists, even if at a low confidence level for $\phi^*$ 
($<2\sigma$), from  $z_m\simeq1.2$ to $z_m\simeq3$ 
where $\phi^*(z\sim3)\simeq0.5\phi^*(z\sim1.2)$ and M$^*$ brightens 
by $\sim0.4$ mag.
It is worth  noting that this evolution  is not dependent on  
the value assumed for $\alpha$.
Indeed, by assuming $\alpha=-1.1$ we obtain for the Ks-band LF fit
M$^*(z\sim0.6)=-24.62$ and $\phi^*(z\sim0.6)=0.0045$ Mpc$^{-3}$
to be compared with M$^*(z\sim1.2)=-25.03$ and $\phi^*(z\sim1.2)=0.0016$ 
Mpc$^{-3}$, resulting in the same luminosity and density evolution (see Tab. 4).

\section{The evolution of the near-IR LF to z$\sim3$}
In this section we compare our results with those obtained by other authors
both at lower and at comparable redshift in order to constrain the evolution 
of the near-IR LF of galaxies from $z\sim3$ to $z\sim0$. 

\subsection{The Ks-band LF}
In Tab. 5 the parameters of the Schechter function of the Ks-band LF
obtained by various authors with different samples are summarized.
In Fig. 19 the fitting Schechter functions obtained by  the various
authors are shown in the range of absolute magnitudes reached
by the  surveys according to the values reported in Tab. 5
(M$_K$(min)).
We first compared the Ks-band LF we derived at $z_m\simeq0.6$ with those 
derived on local samples of galaxies in order to probe and to constrain 
the evolution of the LF on the last 4 Gyrs.
In particular, we considered the LFs of Cole et al. (2001), 
Kochanek et al. (2001) and that of Loveday (2000).
These three local LFs are consistent among them.
The former two  are based on the largest samples of local galaxies
while the latter extends down to very faint luminosities 
(M$_K$(min)$\simeq-16$).
In Fig. 19 the three local LFs are superimposed to the LF we derived at 
$z_m\simeq0.6$ (left upper panel) based on 101 galaxies 50\% of which 
with spectroscopic redshift.
The agreement with the LFs of Kochanek et al. and of Loveday
et al. is rather good.
The largest deviation  is respect to the LF of Cole et al.
However, this deviation is not significant as also confirmed by the K-S 
test which gives P$_{KS}\simeq0.5$. 
The characteristic magnitude of the Schechter
function we derive at $z_m\sim0.6$ are consistent with the local value
(see Tab. 5) even is systematically brighter of about 0.2-0.3 magnitude
while no differences are found in the normalization $\phi^*$. 
Thus, the comparison with the local near-IR LF of galaxies does not point out 
evidence of strong luminosity and/or density evolution  at $z<0.8$.
Given the statistical errors of our estimate we conclude that  the near-IR LF 
has evolved not more than 0.2-0.3 magnitude in the last 4 Gyr.  

\begin{figure*}
\centering
\includegraphics[width=15.cm,height=13.5cm]{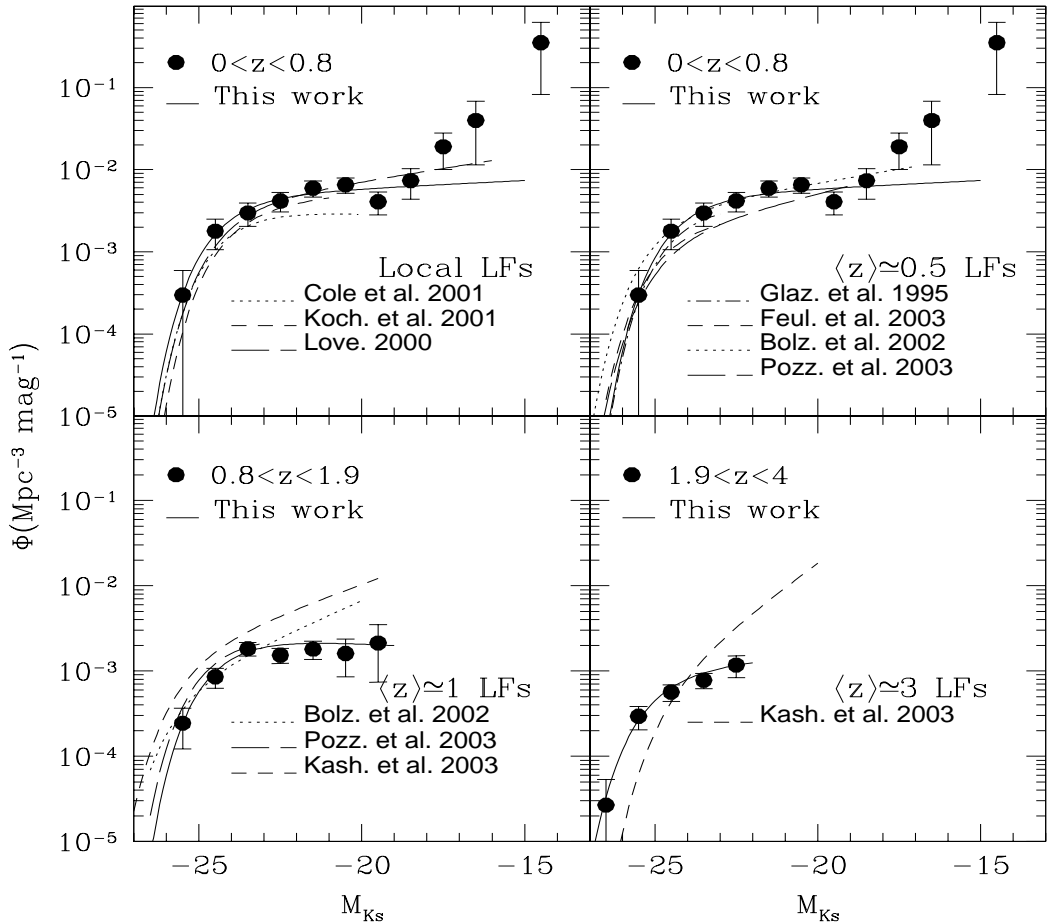}
\caption{The rest-frame Ks-band luminosity function derived with
the 1/V$_{max}$ (points) and  LF Schechter fitting (solid curve)
are compared with the LFs obtained by various authors
(Koch.= Kochaneck; Love.= Loveday; Glaz.= Glazebrook; Feul.= Feulner;
Bolz.= Bolzonella; Pozz.= Pozzetti; Kash.= Kashikawa).
In the upper-left panel the LF we derived in the redshift bin 
$z_{m}\simeq0.6$ is compared with the local K-band LFs.
In the other panels our estimates are compared with those obtained
by other authors at comparable redshift: $z\simeq0.5$ (upper right), 
$z\simeq1$ (lower left) and $z\simeq3$ (lower right).}
\end{figure*}

We then compared our LF at $z_m\sim0.6$ with those derived at comparable 
redshift. 
In the right upper panel of Fig. 19 the Schechter functions representing
the LFs obtained by Glazebrook et al. (1995), Feulner et al. (2003), 
Bolzonella et al. (2002) and Pozzetti et al. (2003)  at $z\sim0.5$
are superimposed to our estimate.
Glazebrook et al. and Feulner et al. derive the LF assuming a 
value of $\alpha$ of -1.0 and -1.1 respectively.
The LF we obtained with these values of $\alpha$, summarized
in Tab. 4, are in  very good agreement with the results of Glazebrook et al. 
and of Feulner et al.
The largest deviations are  respect to the LF of Bolzonella et
al. and of Pozzetti et al.
Both these LFs are characterized by values of $\alpha$ steeper and
by  brighter characteristic magnitudes with respect both to our 
and to the other LFs at these and at lower redshift.
The K-S test we performed shows that the deviation from 
the LF of  Pozzetti et al. is not significant (P$_{KS}\simeq0.15
$) 
while it is significant (P$_{KS}\simeq0.02$) the deviation from 
the LF of Bolzonella et al.
This latter result, rather surprising since we deal with the same field,
is also evident by comparing the LF we obtained at  higher redshift.
In the  lower left panel of Fig. 19 the LFs obtained 
by Pozzetti et al. at $0.75<z<1.3$, by Kashikawa et al. (2003) at $1<z<1.5$
and by Bolzonella et al. at $1<z<2$ are superimposed to the one we obtained in 
the redshift range $0.8\le z<1.9$.
The agreement with the LF of Pozzetti et al. is remarkable.
On the contrary, it is evident that our LF is not consistent with
the LFs of Bolzonella et al. and of Kashikawa et al. who find a very
steep ($\alpha<-1.35$) LF coupled with a very bright (M$^*_K<-25.5$)
characteristic magnitude.
In both the cases the K-S test gives a probability  less than 1\%
that the distributions are drawn from the same parent population.
This is shown in Fig. 20 where the cumulative distributions  relevant
to these LFs  are shown.
The reasons of this disagreement have to be searched both in the different 
samples and magnitudes used by Bolzonella et al. and Kashikawa et al.

As to the LF of Bolzonella et al., even if  relevant to the HDF-S,
it is based on a near-IR sample extracted from an optically selected 
catalogue (Vanzella et al. 2001).
At variance with respect to our sample, 
detection and magnitude estimates were performed and optimized on 
optical HST images.
Moreover, the selection and the completeness of the near-IR sample 
extracted by Bolzonella et al. are based on the 
I$_{814}$-Ks color distribution.
The near-IR data are 4 times shallower  than those used in the present work
and, consequently, the sample is $\sim1$ magnitude shallower.
It should also be noted that 25\% of our sample (50\% of the sample at $z<0.8$)
has spectroscopic redshift and that these redshifts 
have been used both to optimize the photometric redshift
and to estimate the LF. 
All that implies differences in the near-IR magnitude estimates
which can account for the different LF obtained.

As to the comparison with the LF of Kashikawa et al., no obvious reasons can 
be put for the disagreement we obtained in this redshift bin.
Cosmic variance affecting such small areas can affect
the two samples accounting for this discrepancy.  
In the highest redshift bin considered, $z_m\simeq3$, the disagreement is even
larger as shown in the lower-right panel of Fig. 19.
At this redshift, the Ks-band samples rest-frame wavelengths
$\lambda\sim0.55$ $\mu$m and the large extrapolation needed to derive the 
rest-frame Ks-band luminosity can be strongly dependent on the best-fitting template.
In this case,  as  suggested by the analysis we presented in \S 5,
different templates could imply different LFs.
{ However, it should be noted that the disagreement is based on the
comparison of our LF with the Schechter fit found by Kashikawa et al. and
not with their data-points.
The few data-points defining their LF in the two highest redshift bin 
(Fig. 5 of their paper) suggest that the Schechter parameters could 
not be so strongly constrained and that the disagreement could be less severe.}

The agreement with the  Ks-band LFs derived by other authors at 
redshift comparable to $z\sim0.6$ and to $z\sim1.2$ confirms the evolution we 
detect in the LF of galaxies in the HDF-S,  
i.e. a brightening $\Delta M^*\simeq-0.4$ coupled with a decrease of
the number density of bright galaxies 
$\Delta\phi^*/\phi^*\simeq-0.65$ from $z_m\sim0.6$ to $z_m\sim1.2$
for $\alpha=-1.0$.
At larger redshift, the uncertainties in the estimate of rest-frame Ks-band 
luminosities due to the extrapolation based on the best-fitting template could 
affect the LF estimates and mid-IR observations would be useful.
However, even if our LF at $z_m\simeq3$ deviates from the other LFs
at this redshift, these latter LFs trace the same trend we observe 
suggesting that the evolution we detect at $z<2$ extends to $z>3$. 
\begin{figure}
\centering
\includegraphics[width=8.cm,height=7.5cm]{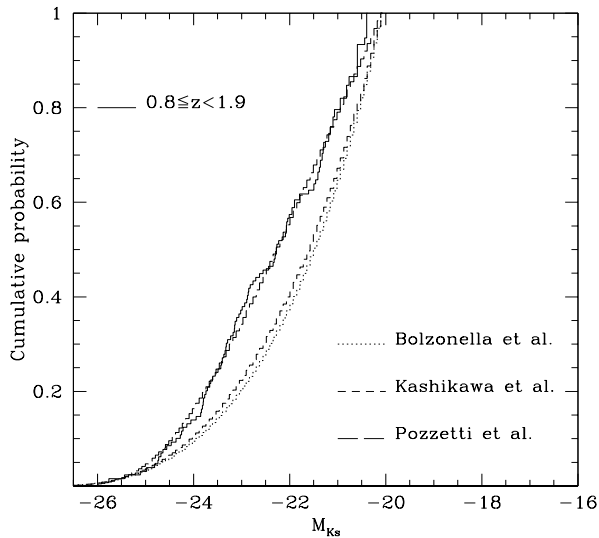}
\caption{Comparison among the cumulative LFs derived at $z\simeq1.2$
normalized over the same range of absolute magnitude. 
The solid histogram is the CLF we derived with the 1/V$_{max}$. 
The probability that the cumulative distributions derived by
the LF of  Bolzonella et al. (dotted line) and of Kashikawa et al.  
(dashed line) are drawn from the population our LF belongs to is
less than 1\%.}
\end{figure}

\subsection{The Js-band LF}
Analogous results are obtained by the comparison of our Js-band LF
estimates with those previously obtained at comparable redshift.
The comparison is shown in Fig. 21 and extends down to $z_m\sim1.2$ since
no Js-band LFs have been estimated at higher redshift by other authors. 
The  local Js-band LF of Cole et al. (2001; $M^*_J=-23.13$ and 
$\phi^*=0.4\cdot 10^{-3}$ Mpc$^{-3}$) and the local LF 
of Balogh et al. (2001, $M^*_J=-23.02$) are consistent with our LF
at $z_m\sim0.6$.
Thus, as in the case of the Ks-band, we do not find evidence
of strong evolution of the LF at $z<0.6$.
However, in agreement with the Ks-band LF, we point out a  
brightening of M$_{Js}^*$ of about 0.2-0.3 magnitude from 
$z=0$ to $z_m\sim0.6$.

The Js-band LFs obtained at $z\sim0.5$ by the various authors show a 
much larger scatter than those obtained in the Ks-band at comparable redshift.
As for the Ks-band, our estimate agrees with the LFs 
of Pozzetti et al. and of Feulner et al. (both based on spectroscopic 
redshift) while deviates significantly from the LFs of Bolzonella et al. 
and of Dahlen et al. (2005).
In particular, both these LFs are characterized by significantly fainter
M$^*$ (0.4 mag with respect to our LF and 0.5-0.6 mag with respect to the
LF of Feulner et al. and Pozzetti et al.).
Also at redshift $z_m\sim 1.2$ our LF agrees very well with the 
Pozzetti et al. LF while deviates significantly from the LFs of 
Bolzonella et al. and Dahlen et al. 
These latter find also an evolution of M$^*$ in the opposite sense
with respect to what we and  other authors found, i.e. a dimming
from $z\sim0.4$ to  $z\sim0.9$.
They use eq. 2 to derive the rest-frame absolute magnitudes.
They suggest that a negative evolution should be expected in Js and 
at longer wavelengths by extrapolating the results obtained by Ilbert et al. 
(2005), who find that the evolution of M$^*$ is systematically weaker
from U to I-bands.
Thus, a turnover in the evolution of M$^*$ could be expected at $\sim$J and a 
negative evolution should take place at longer wavelength.
They suggest that the positive evolution which is instead observed by 
many authors is explained by the method used to derive the rest-frame
luminosities we discussed in \S 4: the observed band (Ks or Js)
samples shorter rest-frame wavelengths (where the evolution is stronger)
at higher redshift and the use of eq. 1 to derive the rest-frame 
luminosities could mimic a brightening in M$^*$.
However, we have shown in \S4 and \S5 that this is not true at least down to
$z\sim2$ and we can consequently exclude with certainty that the evolution 
found by us and by  other authors is due to a wrong derivation of
the rest-frame luminosities.

Thus, from the comparison with the other LFs in Js-band we
confirm an evolution in the LF of galaxies from $z_m\sim0.6$ to $z_m\sim1.2$,  
i.e. a brightening $\Delta M^*\simeq-0.4$ coupled with a decrease of
the number density of bright galaxies.
This evolutionary trend extends to $z\sim3$  even if, as for the 
Ks-band LF, mid-IR observations would be needed to confirm the trend.

\begin{figure}
\centering
\includegraphics[width=9cm,height=11cm]{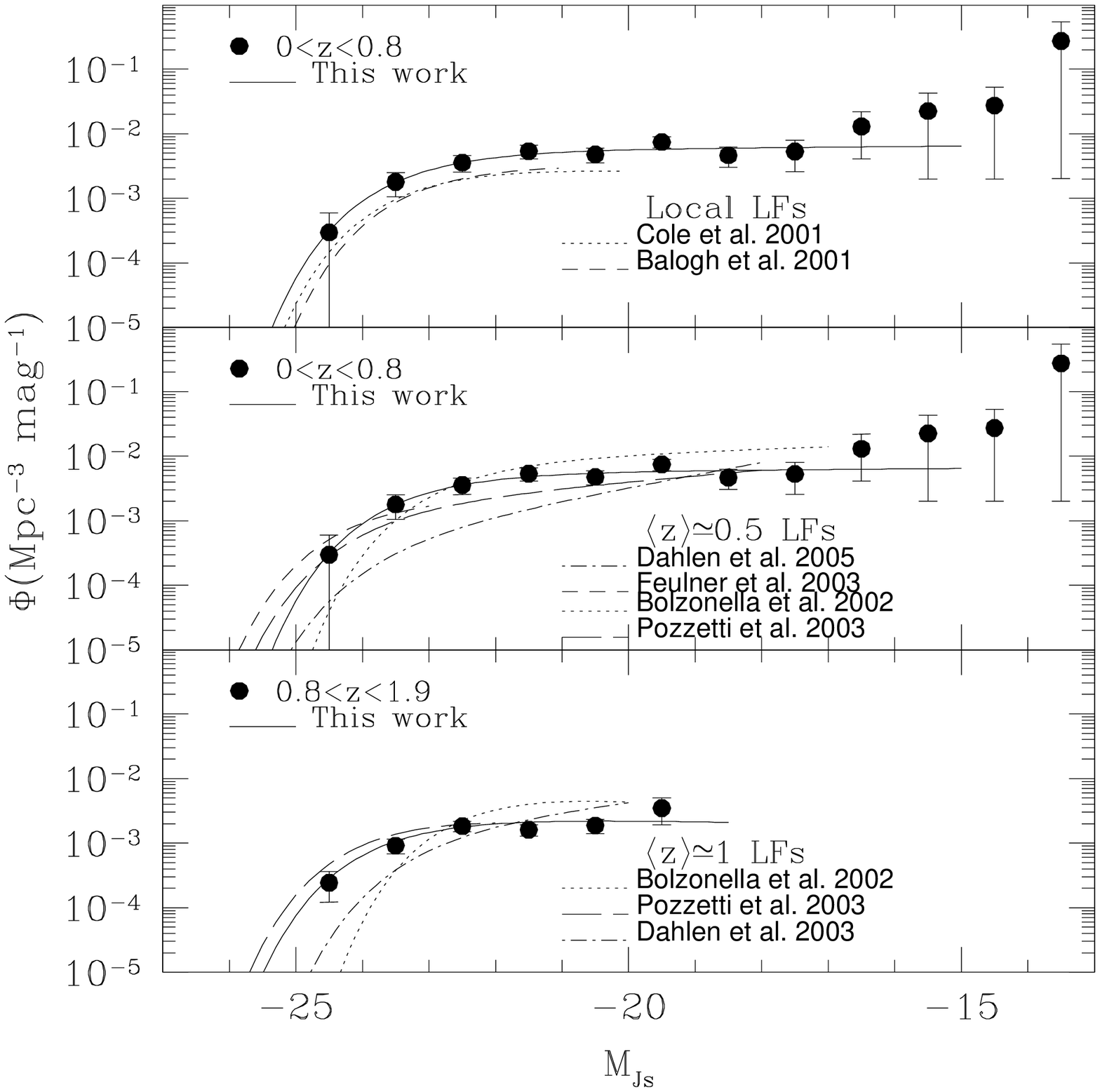}
\caption{The rest-frame Js-band LF derived with
the 1/V$_{max}$ (points) and  LF Schechter fitting (solid curve)
are compared with the LFs obtained by various authors.
In the upper panel the LF  in the redshift bin 
$z_{m}\simeq0.6$ is compared with the local J-band LFs.
In the other panels our estimates are compared with those obtained
by other authors at comparable redshift: $z\simeq0.5$ (middle panel), 
$z\simeq1$ (lower panel)}
\end{figure}

\begin{table*}
\centering
\caption{Summary of the parameters of the Schechter function obtained
by various authors by fitting  the K-band LF of different samples of 
field galaxies. 
For each sample the relevant limiting magnitude, the area, the range of 
redshift covered by the survey (or the mean redshift of the sample) and the 
number of objects are reported. 
The parameters of the Schechter function and the faintest absolute magnitude 
sampled by the data are scaled to the cosmology adopted here ($\Omega_m=0.3$, 
$\Omega_{\Lambda}=0.7$, H$_0=70$ Km s$^{-1}$ Mpc$^{-1}$).}		  
\begin{small}
\begin{tabular}{l c c c c c c c l}
\hline
\hline
Source                  & $m_{lim}$       &              Area &  $z$                 &      obj.s&$M^*$         &$\alpha$       &$\phi^*$   &M$_K$(min)\\
                        &                 &       (arcmin$^2$)&                      &           &              &               &$10^{-3}[Mpc^{-3}]$\\     
\hline
Mobasher et al. 1993\footnotemark[1]    & K$\le$13  B$_J\le$17 &                   &                0-0.1 &        181&$-24.14\pm0.30$&$-1.00\pm0.30$&$3.8\pm0.6$&-22\\
Glazebrook et al. 1995  & K$\le$17.3	  &               552 &                0-0.2 &         55&$-23.91\pm0.23$&$-1.04\pm0.30$&$7.6\pm1.8$&-22\\
                        &		  &                   &                0-0.8 &        124&$-24.42\pm0.11$&$-1.00$       &$4.1\pm0.4$&-23.5\\
Cowie et al. 1996       & K$\le$20 	  &              26.2 &                  0-1 &        393&$-24.27$	 &$-1.25$ &$1.0$	    &-21\\
Gardner et al. 1997     & K$\le$15           &             15840 & $\langle0.14\rangle$ &        465&$-24.07\pm0.17$&$-0.91\pm0.24$&$4.9\pm0.7$&-21.5\\
Szokoly et al. 1998     & K$\le$16.5	  &              2160 &                0-0.4 &        175&$-24.57\pm0.30$&$-1.30\pm0.20$&$3.9\pm1.0$&-21.5\\
Loveday 2000\footnotemark[1]            & b$_{J}\le$17.15   &                   & $\langle0.05\rangle$ &        345&$-24.35\pm0.42$&$-1.16\pm0.19$&$4.1\pm0.3$&-16\\
Kochanek et al. 2001    & K$\le$11.25	  & $<2.5\times 10^7$ & $\langle0.02\rangle$ &       3878&$-24.16\pm0.05$&$-1.09\pm0.06$&$4.0\pm0.3$&-21\\
Cole et al. 2001        & K$\le$13 	  &  $2.2\times 10^6$ & $\langle0.05\rangle$ &      17173&$-24.21\pm0.03$&$-0.96\pm0.05$&$3.7\pm0.6$&-20\\
Feulner et al. 2003     & K$\le$17.5	  &               649 &              0.1-0.3 &        157&$-24.56\pm0.24$&$-1.10$	&$3.8\pm0.4$&-21\\
                        &		  &                   &              0.3-0.6 &        145&$-24.81\pm0.26$&$-1.10$	&$2.4\pm0.9$&-23.5\\
Huang et al. 2003 	& K$\le$15 	  &              1056 & $\langle0.14\rangle$ &           &$-24.47\pm0.08$&$-1.39\pm0.09$&$4.5\pm0.7$&-20.5\\
Pozzetti et al. 2003    & K$\le$20 	  &                52 &             0.2-0.65 &        132&$-24.87\pm0.63$&$-1.25\pm0.22$&$1.8\pm1.2$&-19.5\\
                        &		  &                   &            0.75-1.30 &        170 &$-24.77\pm0.49$&$-0.98\pm0.45$&$2.9\pm1.4$&-23\\
Drory et al. 2003\footnotemark[3]      & K$\le$19.5      & 		  998 &		     0.4-1.2 &$\sim5000$ &               &              &           &-23.5\\
Bolzonella et al. 2002\footnotemark[2] & K$\le$22  	  &                   &                  0-1 &           &$-25.03\pm0.28$&$-1.17\pm0.09$&$3.4\pm1.2$&-17\\
                        &		  &                   &                  1-2 &           &$-25.71\pm1.14$&$-1.42\pm0.10$&$0.8\pm0.7$&-20.5\\
Caputi et al. 2005\footnotemark[3]    &K$\le$22		&		  50  &0-2.5   &		1600&		&		&		& \\
Kashikawa et al. 2003\footnotemark[3]  & K$\le$24 	  &           $\sim4$      &                0.6-1 &   439        &$-25.41\pm0.20$&$-1.35\pm0.04$&$2.3\pm0.7$&-18.5\\
                        &		  &                   &                1-1.5 &           &$-25.53\pm0.27$&$-1.35\pm0.06$&$1.9\pm0.8$&-19.5\\
                        &		  &                   &              1.5-2.5 &           &$-25.58\pm0.12$&$-1.37\pm0.05$&$1.3\pm0.5$&-19.5\\
                        &		  &                   &              2.5-3.5 &           &$-24.66\pm0.23$&$-1.70\pm0.11$&$1.0\pm0.7$&-20\\
This work\footnotemark[3]             & K$\le$23 	  &               5.5 &                0-0.8 &        101&$-24.47\pm0.63$&$-1.14\pm0.11$&$4.0\pm2.0$&-14\\
                        &		  &                   &              0.8-1.9 &        100&$-24.53\pm0.37$&$-0.85\pm0.17$&$2.9\pm1.0$&-19\\
                        &		  &                   &                1.9-4 &         84&$-25.18\pm0.53$&$-1.10\pm0.31$&$1.0\pm0.6$&-22\\
\hline					  
\end{tabular}
\end{small}				  
\footnotesize
\flushleft
\footnotemark[1] Based on K-band imaging of an optically selected sample

\footnotemark[2] Based on K-band data of an optically selected sample and on photometric redshift.

\footnotemark[3] Based on photometric redshift. The number of galaxies (439) is 
the total number of galaxies in the sample.
\normalsize
\end{table*}

\section{Summary and conclusions}
We have probed the evolution of the rest-frame Js-band and Ks-band 
LFs of field galaxies with a complete sample of about 300 galaxies selected 
in the HDF-S at Ks$\le23$ (Vega).
Photometric redshifts have been obtained from template SED fitting to 
U$_{300}$, B$_{450}$, V$_{606}$, I$_{814}$, Js, H and Ks photometry 
after having calibrated
and optimized the procedure and the set of templates with a control sample of 
232 spectroscopic redshifts, 151 from the HDF-N and 81 from the HDF-S. 
The accuracy in the redshift estimate we obtained is 0.06 (rms). 

We investigate the reliability of the rest-frame near-IR 
absolute magnitudes obtained using the conventional method based on
the extrapolation of the observed photometry on the best-fitting template
and  using the photometry approaching the rest-frame near-IR.
We find that the rest-frame Js-band absolute magnitudes
obtained through the photometry in the redder bands (H and Ks
according to the redshift of the galaxy)
are consistent with those obtained from the Js-band photometry
at least down to $z\sim2$.
This shows that the LF is not dependent either on the extrapolation made
on the best-fitting template or on the library of models used.

We derived the Js-band and the Ks-band LF in the three redshift bins
[0,0.8), [0.8,1.9) and [1.9,4) centered at the median redshift
$z_m\sim0.6$, $z_m\sim1.2$ and $z_m\sim3$.
Each bin contains 101, 100 and 84 galaxies of which 52 (50\%),
12 (12\%) and 10 (12\%) with spectroscopic redshift respectively. 
The analysis of the observed LF at different redshifts and the
comparison with those previously found by other authors at comparable 
redshift and at $z=0$ provided the following results:

\noindent
1. We find hints of a raise of the faint end (M$_{Js}>-17$ and 
M$_{Ks}>-18$) of the near-IR LF at $z_m\sim0.6$.
The raise is defined by 9 galaxies at $z<0.3$ with irregular morphology 
as found for the local LF at optical wavelength (e.g. Marzke et al. 1994;  
Marzke et al. 1998; Zucca et al. 1997; Folkes et al. 1999).
They account for a comoving number density $\bar n=0.4\pm0.23$ Mpc$^{-3}$
at $<z>\simeq0.2$, almost one order of magnitude higher than that
of brighter galaxies. 
{ However, given the low statistics, such raise of the
faint end  needs a larger sample to be established;}

\noindent
2. We find no evidence of a steepening with redshift of the near-IR LFs 
of galaxies in the HDF-S.
The value of $\alpha$ we find is consistent with the local value. 
Given the size of our sample, we estimate that if $\alpha$ changes with 
redshift its evolution is constrained within $\pm0.3$ from $z\sim0$ to 
$z\sim3$.

\noindent
3. We do not find evidence of strong evolution of the LF up to $z_m\simeq0.6$,
where 50\% of our galaxies has spectroscopic redshift.
The comparison with the local estimates of the near-IR LF (Cole et al. 2001;
Kochanek et al. 2001) shows that
M$^*$ has evolved not more than 0.2-0.3 magnitude in this redshift range,
while the number density of galaxies ($\phi^*$) has not evolved
significantly in this redshift range, in agreement with the previous studies 
(e.g. Drory et al. 2003; Pozzetti et al. 2003; Feulner et al. 2003; 
Glazebrook et al. 1995; Caputi et al. 2004, 2005).

\noindent
4. We clearly detect  an evolution of the LF at  $z_m>0.6$ characterized by a 
brightening of M$^*$ of about 0.6 and a decrease of $\phi^*$ by a factor 
2-3 to $z_m\sim1.2$ both in the Js-band and in the Ks-band. 
{ By fixing $\alpha=1.0$, this evolution  is characterized by 
$\Delta M^*\simeq-0.6$ (at 1$\sigma$) coupled with 
$\Delta\phi^*/\phi^*\simeq-0.65$ (at $>3\sigma$).
The brightening  persists (at 1$\sigma$) up to $z_m\sim3$
together with the decline of $\phi^*$ (at $\sim 2\sigma$) 
even if at a lower extent.}

Our results agrees, at least qualitatively, with most of the analysis 
previously done (Drory et al. 2003; Pozzetti et al. 2003; Feulner et al. 2003;
Caputi et al. 2004, 2005) while deviates from the recent estimate of 
Dahlen et al. (2005).  
Taking into account that the near-IR light is rather tightly connected to
the stellar mass, our results can  give insights on the evolution
of the stellar mass of the galaxies besides their luminosity. 
Up to $z\simeq0.8$, clearly little luminosity evolution and no
density evolution is observed.
This suggests that the population of local bright galaxies was 
already formed at $z\sim0.8$ and that their stellar mass growth was
already completed at this redshift. 
This agrees with the results on the evolution of the stellar 
mass density obtained by many authors: Fontana et al. (2003) and
Rudnick et al. (2003) in  the HDF-S,
Fontana et al. (2004) on the K20 sample (Cimatti et al. 2002), 
Yamada et al. (2005) in the Subaru Deep Survey Field,
 Bundy et al. (2005) and Drory et al. (2005)
in the GOODS and FDF fields and  Feulner et al. (2005a) in the MUNICS fields.
Indeed, all of them  find very little or even no evolution in the
stellar mass function of galaxies and in the specific star formation 
rate (SSFR) at $z<1$ for high-mass galaxies.

At $z>0.8$ the evolution is stronger and the larger brightening observed
is accompanied by a decrease of the number density of bright/massive galaxies.
In the redshift bin [0.8,1.9) the number density of bright galaxies is
$\sim30-50$\% of the local value and reaches $\sim20-30$\% at $1.9<z<4$.
This decline in the number density   
suggests that up to $70$\%  of the local massive/bright galaxies has grown 
at $1<z<2-3$ through star formation, merging or both.
Thus, we should expect to observe massive 
star-forming galaxies with disturbed/irregular morphology and merging systems
in this high redshift range, as indeed found in same cases 
(e.g. Daddi et al. 2004).
On the other hand, the observed evolution implies also that at least 30\% 
of the local bright/massive galaxies was already 
in place at $1<z<2$  as indeed found by some authors
(Fontana et al. 2004; Glazebrook et al. 2004; F$\ddot{\rm o}$rster Schreiber et al.
2004).
Pozzetti et al. (2003) and Caputi et al. (2005a) show that the bright
end of the LF is dominated by early-type galaxies and that their 
contribution to the bright end does not decline with redshift.
Thus, that 30\% should be dominated by early-types implying that we should 
observe  massive evolved galaxies fully 
assembled at $1<z<2$  and even at $z>2$.
This is indeed observed by many authors (Cimatti et al. 2004; 
McCarthy et al. 2004; Saracco et al. 2004, 2005; 
Longhetti et al. 2005; Daddi et al. 2005; Labb\'e et al. 2005).  
It is worth noting that this picture could be accounted for, at least 
qualitatively, 
in the hierarchical picture of galaxy formation as suggested by the recent 
results obtained by Nagamine et al. (2005).

From these considerations we can gather that the growth of massive galaxies 
does not follow a unique way but displays different behaviours.
A significant fraction (50-70\%) of the brighetst/most massive galaxies
increases their stellar mass 
over a large redshift range  at $z>1$.
The remaining fraction reaches their final mass in a narrower redshift range
at $z>3$ since it is already in place by this redshift.
This suggests that, for the former, the stellar mass 
growth has been less efficient and has proceeded more slowly than for the 
latter. 
On the contrary, in the latter, the stellar mass has to 
be grown rapidly in a short interval, surely much shorter than 1 Gyr 
given their redshift.
This is  supported by the recent results derived by mid-IR  
massive galaxies observed at high-z (e.g. Caputi et al. 2005b)
and suggests a high efficiency in the accretion of the
stellar mass in massive haloes in the early Universe, possibly through
a very efficient star formation.
The recent results obtained on the evolution of the SSFR of galaxies 
seems to  point toward this direction.
The strong increase in the SSFR of the most massive galaxies with redshift
(Feulner et al. 2005b) favors indeed an efficient SFR in 
the brightest galaxies at $z>3-4$ constraining their growth in a very short 
interval.

\section*{Acknowledgments}
This work is based on observations made with the ESO-VLT telescopes at 
Paranal Observatory under the programs 164.O-0612 and 70.B-0144 and
with the NASA/ESA Hubble Space Telescope. 
We are in debt with M. Bolzonella for her helpfulnes. 
We thank the anonymous referee for the useful comments which improved 
the presentation of the results.

\label{lastpage}


\begin{thebibliography}{}
\bibitem[]{} Balogh M. L., Christlein D., Zabludoff A. I., Zaritsky D., 2001,
ApJ, 557, 117
\bibitem[]{} Bell E. F., McIntosh D. H., Katz N., Weinberg M. D., 2003, ApJS, 149, 289
\bibitem[]{} Bell E. F., et al. 2004, ApJ, 608, 752
\bibitem[]{} Bertin E., Arnouts S., 1996, A\&AS, 117, 393
\bibitem[]{} Blanton M. R., et al., 2003, ApJ, 592, 819
\bibitem[]{} Bolzonella M., Miralles J.-M., Pello\' R., 2000, A\&A, 363, 476
\bibitem[]{} Bolzonella M., Pell\`o R., Maccagni D. 2002, A\&A, 395, 443
\bibitem[]{} Bruzual A.,G. \& Charlot S. 2003, MNRAS, 344, 1000
\bibitem[]{} Bundy K., Ellis R. S., Conselice C. J., 2005, ApJ, 625, 621
\bibitem[]{} Calzetti D., Armus L., Bohlin R. C., Kinney A. L., Koornneef J., Storchi-Bergmann T., 2000, ApJ, 533, 682
\bibitem[]{} Caputi K. I., Dunlop J. S., McLure R. J., Roche N. D., 2004, 
MNRAS, 353, 30
\bibitem[]{} Caputi K. I.,  Dunlop J. S., McLure R. J., Roche N. D., 2005a, 
MNRAS, 361, 607
\bibitem[]{} Caputi K. I.,  Dole H., Lagache G., et al., 2005b, ApJ, in press, [astro-ph/0510070]
\bibitem[]{} Casertano S., et al., 2000, AJ, 120, 2747
\bibitem[]{} Cimatti A., Mignoli M., Daddi E., et al. 2002, A\&A, 392, 395
\bibitem[]{} Cimatti A., et al., 2004, Nat., 430, 184
\bibitem[]{} Cohen J. G., Hogg D. W., Blandford R., et al. 2000, ApJ, 538, 29
\bibitem[]{} Cole S., Norberg P., Baugh C. M., et al. 2001, MNRAS, 326, 255
\bibitem[]{} Coleman G. D., Wu C.-C., Weedman D. W. 1980, ApJS, 43, 393
\bibitem[]{} Cowie L. L., Songaila A., Hu E. M., Cohen J. G., 1996, AJ, 112, 839
\bibitem[]{} Daddi E., Cimatti A., Renzini A., et al., 2004, ApJ, 600, L127
\bibitem[]{} Daddi E., et al., 2005, ApJ, 626, 680
\bibitem[]{} Dahlen T., Mobasher B., Somerville R. S., Moustakas L. A., 
Dickinson M., Ferguson H. C., Giavalisco M., 2005, ApJ, 631, 126
\bibitem[]{} Dawson S., Stern D., Bunker A. J., Spinrad H., Dey A., 2001, 
AJ, 122, 598
\bibitem[]{} Drory N., Bender R., Feulner G., Hopp U., Maraston C., 
Snigula J., Hill G. J. 2003, ApJ, 595, 698 
\bibitem[]{} Drory N., Salvato M., Gabash A., Bender R., Hopp U., Feulner G., Pannella M., 2005, ApJ, 619, L131
\bibitem[]{} Fern\'andez-Soto A., Lanzetta K. M., Chen H.-W., Levine B., Yahata N., 2002, MNRAS, 330, 889
\bibitem[]{} Feulner G., Bender R., Drory N., Hopp U., Snigula J., Hill G. J.,
2003, MNRAS, 342, 605
\bibitem[]{} Feulner G., Goranova Y., Drory N., Hopp U., Bender R., 2005a, MNRAS, 358, L1
\bibitem[]{} Feulner G., Gabasch A., Salvato M., Drory N., Hopp U., Bender R., 2005b, ApJ, 633, L9 
\bibitem[]{} Fioc M., Rocca-Volmerange B., 1997, A\&A, 326, 950
\bibitem[]{} Folkes S., Ronen S., Price I., et al., 1999, MNRAS, 308, 459
\bibitem[]{} Fontana A., Donnarumma I., Vanzella E., et al. 2003, ApJ, 594, L9
\bibitem[]{} Fontana A., Pozzetti L., Donnarumma I., et al. 2004, A\&A 424, 23
\bibitem[]{} F$\ddot{\rm o}$rster Schreiber N. M., et al. 2004, 616, 40
\bibitem[]{} Franx M., et al., 2000, Messenger, 99, 20
\bibitem[]{} Gabasch A., et al., 2004, A\&A, 421, 41
\bibitem[]{} Gardner J. P., Sharples R. M., Frenk C. S., Carrasco B. E., 1997,
ApJ, 480, L99
\bibitem[]{} Giallongo et al. 2005, ApJ, 622, 116
\bibitem[]{} Glazebrook K., Peacock J. A., Miller L., Collins C. A., 1995, MNRAS, 275, 169
\bibitem[]{} Glazebrook K., Abraham R. G., McCarthy P. J., et al., 2004, Nat., 430, 181
\bibitem[]{} Huang J.-S., Glazebrook K., Cowie L. L., Tinney C., 2003, ApJ, 584, 203
\bibitem[]{} Ilbert O., et al. 2005, A\&A, 439, 863
\bibitem[]{} Kashikawa N., et al., 2003, AJ, 125, 53
\bibitem[]{} Kochanek C. S., Pahre M. A., Falco E. E., et al. 2001, 
ApJ, 560, 566
\bibitem[]{} Labb\`e I., et al. 2003a, AJ, 125, 1107
\bibitem[]{} Labb\`e I., et al. 2003b, ApJ, 591, L95
\bibitem[]{} Labb\`e I., et al. 2005, ApJ, 624, L81
\bibitem[]{} Lilly S. J., Tresse L., Hammer F., Crampton D., Le Fevre O. 1995,
ApJ, 455, 108
\bibitem[]{} Lilly S. J., Le Fevre O., Crampton D., Hammer F., Tresse L., 
1995b, ApJ, 455, 50
\bibitem[]{} Lin H., Yee H. K. C., Carlberg R. G., Ellingson E., 1997, ApJ, 475, 494
\bibitem[]{} Liu C. T., Green R. F., Hall P. B., Osmer P. S., 1998, AJ, 116, 1082
\bibitem[]{} Longhetti M., Saracco P., Severgnini P., et al., 2005, MNRAS, 361, 897
\bibitem[]{} Loveday J., 2000, MNRAS, 312, 557
\bibitem[]{} Madgwick D. S., et al., 2002, MNRAS, 333, 133
\bibitem[]{} Mannucci F., Basile F., Poggianti B. M., Cimatti A., Daddi E., Pozzetti L., Vanzi L., 2001, MNRAS, 326, 745
\bibitem[]{} Maraston C., 2005, MNRAS, 362, 799
\bibitem[]{} Marzke R. O., Geller M., J., Huchra J., Corwin H., G. Jr., 1994
AJ, 108, 437
\bibitem[]{} Marzke R. O., da Costa L. N., Pellegrini P. S., Willmer C. N. A., 
Geller M., J., 1998, ApJ, 503, 617
\bibitem[]{} McCarthy P. J., Le Borgne D., Crampton D., et al. 2004, ApJ, 614, L9
\bibitem[]{} Mobasher B., Sharples R. M., Ellis R. S., 1993, MNRAS, 263, 560
\bibitem[]{} Nagamine K., Cen R., Hernquist L., Ostriker J. P., Springer F., 2005, ApJ, 627, 608
\bibitem[]{} Norberg P., et al., 2002, MNRAS, 336, 907
\bibitem[]{} Poli F., et al., 2003, ApJ, 593, L1
\bibitem[]{} Pozzetti L., Cimatti A., Zamorani G., et al., 2003, A\&A, 402, 837
\bibitem[]{} Pr\'evot M. L., Lequeux J., Pr\'evot L., Maurice E., Rocca-Volmerange B., 1984, A\&A, 132, 389
\bibitem[]{} Rigopoulou D., Vacca W. D., Berta S., Franceschini A., 
Aussel H., 2005, A\&A, 440, 61
\bibitem[]{} Rix H., Riecke M. J., 1993, ApJ, 418, 123
\bibitem[]{} Rudnick G., Rix H.-W., Franx M., et al. 2003, ApJ, 599, 847
\bibitem[]{} Saracco P., Giallongo E., Cristiani S., et al., 2001, A\&A, 375, 1
\bibitem[]{} Saracco P., Longhetti M., Giallongo E., et al. 2004, A\&A, 420, 
125
\bibitem[]{} Saracco P., Longhetti M., Severgnini P., et al., 2005, MNRAS, 
357, L40
\bibitem[]{} Sawicki M., Mall\'en-Ornelas G., 2003, AJ, 126, 1208
\bibitem[]{} Schmidt M. 1968, ApJ, 151, 193
\bibitem[]{} Szokoly G. P., Subbarao M. U., Connolly A. J., Mobasher B., 1998, ApJ, 492, 452
\bibitem[]{} Takeuchi T. T., Yoshikawa K., Ishii T. T. 2000, ApJS, 129, 1
\bibitem[]{} Trujillo I., et al., 2004, ApJ, 604, 521
\bibitem[]{} Vanzella E., et al., 2002, A\&A, 396, 847
\bibitem[]{} Vanzella E., et al., 2001, AJ, 122, 2190
\bibitem[]{} Wolf C., Meisenheimer K., Rix H. W., Borch A., Dye S., 
Kleinheinrich M., 2003, A\&A, 401, 73
\bibitem[]{} Yamada T., et al., 2005, ApJ, 634, 861
\bibitem[]{} Zucca E., Zamorani G., Vettolani G., et al., 1997, A\&A, 326, 477
\end{thebibliography}
\end{document}